\documentclass[aps,prd,twocolumn,superscriptaddress,preprintnumbers,floatfix,nofootinbib]{revtex4-1}
\pdfoutput = 1
\usepackage{amssymb}
\usepackage{amsmath}
\usepackage{epsfig}
\usepackage[usenames,dvipsnames]{color}
\usepackage{hyperref}
\usepackage{comment}
\usepackage{color}
\usepackage{cleveref}
\usepackage{multirow}
\usepackage{capt-of}
\usepackage{siunitx}
\usepackage{graphicx}
\usepackage{gensymb}
\usepackage[caption=false]{subfig}
\usepackage{slashed}
\usepackage{tabularx}
\usepackage{enumitem}
\usepackage{multirow}
\usepackage[normalem]{ulem}
\usepackage{aas_macros}

\hypersetup{
    colorlinks=true,       
    linkcolor=Cyan,        
    citecolor=Magenta
    }

\makeatletter

\begin{document}

\title{Soliton Merger Rates and Enhanced Axion Dark Matter Decay}

\author{Xiaolong Du}
\email{xdu@astro.ucla.edu}\thanks{-- \href{https://orcid.org/0000-0003-0728-2533}{0000-0003-0728-2533}}
\affiliation{Carnegie Observatories, 813 Santa Barbara Street, Pasadena, CA 91101, USA}
\affiliation{Department of Physics and Astronomy, University of California, Los Angeles, CA 90095, USA}

\author{David J. E. Marsh}
\email{david.j.marsh@kcl.ac.uk}\thanks{-- \href{https://orcid.org/0000-0002-4690-3016}{0000-0002-4690-3016}}
\affiliation{Theoretical Particle Physics and Cosmology, King's College London, Strand, London, WC2R 2LS, United Kingdom}

\author{Miguel Escudero}
\email{miguel.escudero@cern.ch}\thanks{-- \href{https://orcid.org/0000-0002-4487-8742}{0000-0002-4487-8742}}
\affiliation{Theoretical Physics Department, CERN, 1211 Geneva 23, Switzerland}

\author{Andrew Benson}
\email{abenson@carnegiescience.edu}\thanks{-- \href{https://orcid.org/0000-0001-5501-6008}{0000-0001-5501-6008}}
\affiliation{Carnegie Observatories, 813 Santa Barbara Street, Pasadena, CA 91101, USA}

\author{Diego Blas}
\email{blas.diego@gmail.com}\thanks{-- \href{https://orcid.org/0000-0003-2646-0112}{0000-0003-2646-0112}}
\affiliation{Grup de F\'isica Te\`{o}rica, Departament de F\'isica, Universitat Aut\`{o}noma de Barcelona, 08193 Bellaterra (Barcelona), Spain}
\affiliation{Institut de F\'isica d’Altes Energies (IFAE), The Barcelona Institute of Science and Technology, Campus UAB, 08193 Bellaterra (Barcelona), Spain}

\author{Charis Kaur Pooni}
\email{charis.pooni@kcl.ac.uk}\thanks{-- \href{https://orcid.org/0000-0002-6658-3478}{0000-0002-6658-3478}}
\affiliation{Theoretical Particle Physics and Cosmology, King's College London, Strand, London, WC2R 2LS, United Kingdom}

\author{Malcolm Fairbairn}
\email{malcolm.fairbairn@kcl.ac.uk}\thanks{-- \href{https://orcid.org/0000-0002-0566-4127}{0000-0002-0566-4127}}
\affiliation{Theoretical Particle Physics and Cosmology, King's College London, Strand, London, WC2R 2LS, United Kingdom}

\begin{abstract}
Solitons are observed to form in simulations of dark matter (DM) halos consisting of bosonic fields. We use the extended Press-Schechter formalism to compute the mass function of solitons, assuming various forms for the relationship between halo mass and soliton mass. We further provide a new calculation of the rate of soliton major mergers. Solitons composed of axion DM are unstable above a critical mass, and decay to either relativistic axions or photons, depending on the values of the coupling constants. We use the computed soliton major merger rate to predict the enhanced DM decay rate due to soliton instability. For certain values of currently allowed axion parameters, the energy injection into the intergalactic medium from soliton decays to photons is comparable to or larger than the energy injection due to core collapse supernovae at $z>10$. A companion paper explores the phenomenology of such an energy injection.
\end{abstract}

\preprint{KCL-PH-TH-2023-03, CERN-TH-2023-009}

\maketitle

\section{Introduction}\label{sec:intro}

Real scalar fields in general relativity have time-periodic, spatially localized, finite energy ground state solutions~\cite{Seidel:1991zh,Seidel:1993zk}. These solutions are known by various names: oscillatons, solitons, axion stars, and so forth. The inward force of gravity is balanced in these configurations by the outward pressure of scalar field gradients. In the nonrelativistic limit, these objects are truly stationary ground state soliton solutions of the Schr\"{o}dinger-Poisson equations (see, e.g., Refs.~\cite{Guzman:2006yc,Chavanis:2011zi,Marsh:2015xka,Hui:2016ltb}).~\footnote{Solitons also form for massive spin-1 fields~(see, e.g., Refs.~\cite{Jain:2021pnk,Gorghetto:2022sue} and references therein). In the following we simplify the discussion to scalar fields, although our methods can also be applied more generally.} 

If the observed~\cite{aghanim2020planck} cosmological dark matter (DM) is composed of a real scalar field, or scalar fields, then such solitons should form in our Universe. Indeed, in numerical simulations of both axion~\cite{Eggemeier:2019jsu} and ``fuzzy'' DM~\cite{Schive:2014dra} solitons form in the centers of DM halos during the earliest stages of collapse, driven by the initial coherence of the field on scales near the de Broglie wavelength. Furthermore, solitons can form by gravitational Bose-Einstein condensation~\cite{Levkov:2018kau,Chen:2020cef} in any environment where the condensation timescale is shorter than the age of the Universe.

The possible existence of solitons, being very dense and coherent lumps of DM, opens a wide range of possibilities concerning their phenomenology. Soliton cores can affect stellar motions in the centers of galaxies (e.g., Refs.~\cite{Marsh2015b,Gonzalez-Morales:2016yaf,Bar:2018acw,2021ApJ...916...27D}). As ``exotic compact objects," gravitational waves (GWs) from soliton mergers can appear distinctly in GW observations (e.g., Ref.~\cite{Giudice:2016zpa}). Soliton instabilities can also lead to new formation channels for black holes~\cite{Helfer:2016ljl,Widdicombe:2018oeo}, or production of relativistic particles~\cite{Levkov:2016rkk,Levkov:2020txo}.

Exploiting solitons as a phenomenological window onto DM requires knowing the cosmological distribution of solitons, and their merger rates, which we dedicate this paper to studying. We study two mechanisms by which solitons lead to enhanced DM decay: by plasma blocking of parametric resonance, and by major mergers leading to formation of supercritical solitons.   

In the present work, we will be concerned with soliton cores formed at very early times, $z\gtrsim 10$, in some of the first DM halos with masses $M\gtrsim 10^{-5}M_\odot$. The rough particle mass scale we are concerned with is $m_a\approx 10^{-11}\text{ eV}$. For this mass scale, assuming an axionlike particle (ALP) with temperature-independent mass, halo formation is strongly suppressed for $M<8\times 10^{-5}M_\odot$~\cite{Schive:2015kza}. Thus, the formation of the halos of interest is expected to be very similar to the formation of the first halos in thermal WIMP models, where the minimum halo mass is $M\approx 10^{-6}M_\odot$~\cite{Green:2005fa,Loeb:2005pm,Angulo:2016qof}. For standard $\Lambda$CDM cosmology the power spectrum on small scales is close to a fixed power law, which, together with the scaling symmetries of the Schr\"{o}dinger-Poisson equations (and ignoring baryonic feedback), implies that the halos of interest formed within a few orders of magnitude of the cutoff scale should be morphologically similar to the $\mathcal{O}(10^{10}M_\odot)$ halos formed at $z\approx 8$ and simulated directly for $m_a\approx 10^{-22}\text{ eV}$~\cite{Schive:2014dra}. Crucially, these first formed halos are expected to host a single soliton at their center and to obey a ``core-halo mass relation''~\cite{Schive:2014hza}, which we discuss in detail below.

\begin{figure}[t]
\begin{center}
\includegraphics[width=\columnwidth]{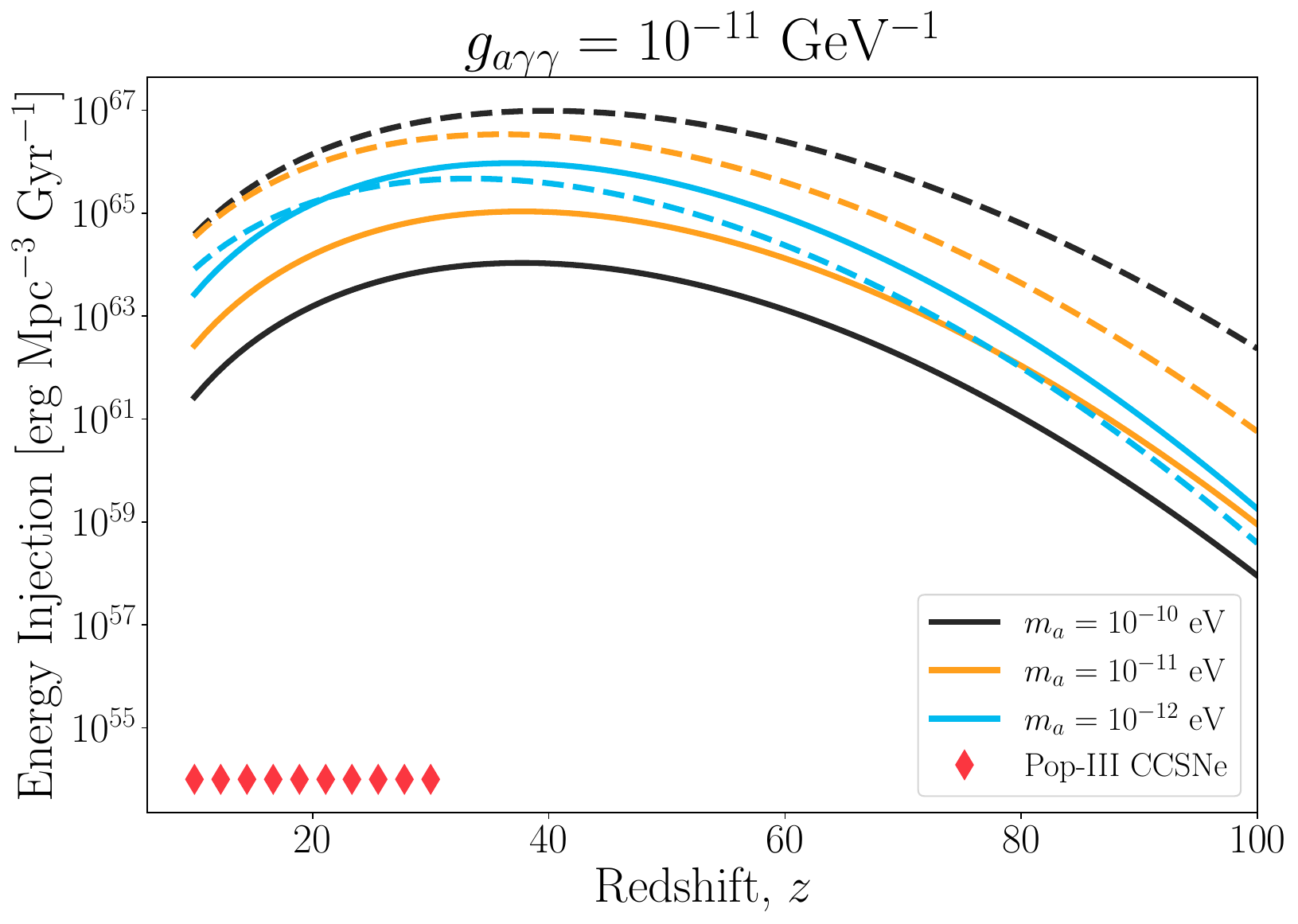}
\caption{Energy injection into the intergalactic medium from axion dark matter decay caused by soliton mergers. For comparison, a typical energy scale and redshift for energy injection from Pop-III core-collapse supernovae (CCSNe) is indicated~\cite{Hartwig:2022lon}. For the parameters shown, axion star explosions inject more energy than supernovae. See~\cite{Escudero:2023vgv} for the cosmological implications of such decays.Solid and dashed lines indicate different assumptions for the core-halo mass relation of $\alpha=1/3$ and $\alpha=3/5$.}
\label{fig:energy}
\end{center}
\end{figure}

A key result of the present work is the computation of the energy injection into the intergalactic medium caused by the decay of axion dark matter due to soliton mergers. Therefore, note that in most of this work we use the terms ``soliton" and ``axion star" interchangeably. The energy injection density is shown for some representative parameters in Fig.~\ref{fig:energy}. We compare this to an approximate energy injection due to core collapse supernovae from Pop-III stars~\cite{Hartwig:2022lon}. The energy injection from axion dark matter soliton decay exceeds the supernova energy by many orders of magnitude, and extends to much higher redshifts, deep into the dark ages. This suggests that this phenomenon can place new constraints on axion dark matter, and open a new window onto axion observation. Details are explored in the companion paper~\cite{Escudero:2023vgv}.

This paper is organized as follows. In Sec.~\ref{sec:mass_func} we compute the soliton mass function assuming various models for the core-halo mass relation between DM halos and soliton masses. We discuss the critical soliton mass in Sec.~\ref{sec:critical-soliton}, and compute the DM fraction that decays instantaneously at a given redshift due to plasma blocking. In Sec.~\ref{sec:merger_rate} we use the extended Press-Schechter model and stochastic merger trees to compute the soliton merger rate, and the DM decay rate from the formation of super-critical solitons by major mergers of sub-critical solitons. We conclude in Sec.~\ref{sec:conclusions}. The Appendices give details of our numerical scheme. We adopt fixed cosmological parameters $\Omega_M=0.3153$, $\Omega_b=0.04930$, $h=0.6736$, $\sigma_8=0.8111$, $n_s=0.9649$~\cite{Planck:2018jri}.

\section{Soliton Mass Function}\label{sec:mass_func}

We consider DM composed of a real scalar field $\phi$ minimally coupled to gravity with canonical kinetic term and potential $V(\phi)=m_a^2\phi^2/2+\lambda\phi^4/4!$, with $m_a$ the DM particle mass (axion mass). In the nonrelativistic limit, solitons are given by the ground-state solutions of the Schr\"{o}dinger-Poisson equations:
\begin{align}
    i\partial_t \psi &=-\frac{1}{2m_a}\nabla^2\psi+m_a\Phi \psi+\lambda |\psi|^2\psi, 
    \label{eq:schroedinger} \\
    \nabla^2 \Phi &= 4\pi G |\psi|^2\, ,
    \label{eq:poisson}
\end{align}
where $\Phi$ is the Newtonian gravitational potential, $\lambda$ is the self-interaction coupling, and we use units $\hbar=c=1$. The field $\psi$ is related to the fundamental scalar field $\phi$ by the WKB approximation:
\begin{equation}
    \phi = \frac{1}{\sqrt{2}m_a}\left( e^{im_at}\psi + e^{-im_a t}\psi^*\right)\, .
    \label{eq:WKB}
\end{equation}
In the relativistic limit, solitons are found as the time periodic solutions of the Einstein-Klein-Gordon equations~\cite{Seidel:1991zh,Seidel:1993zk}. The soliton solutions possess a scaling symmetry, and as such are uniquely specified by the soliton mass, or alternatively the central field value, $\phi_0$.

\begin{figure*}[t]
\begin{center}
\includegraphics[width=\columnwidth]{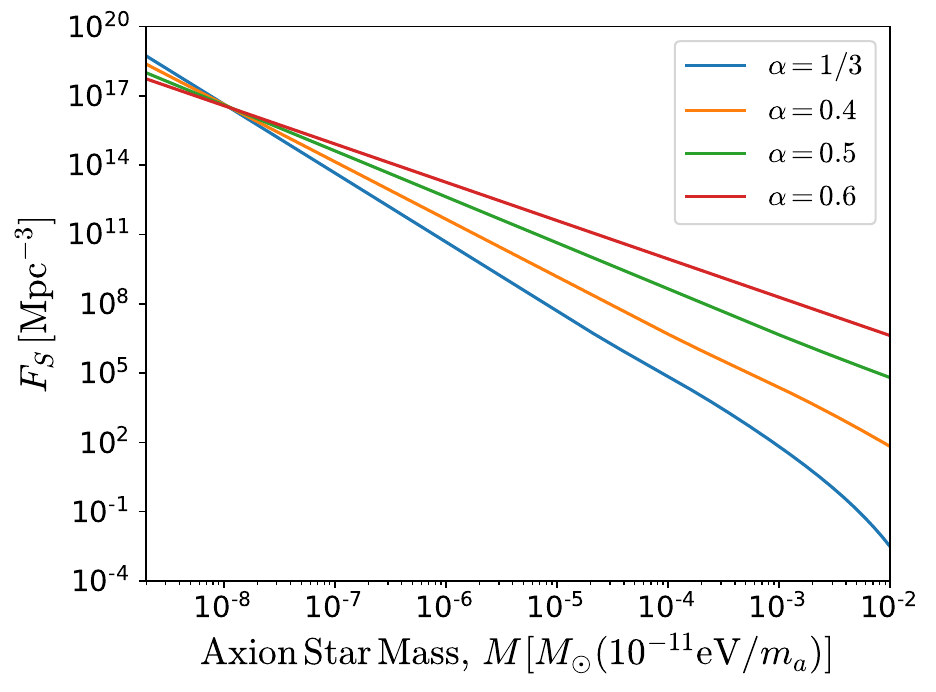}
\includegraphics[width=\columnwidth]{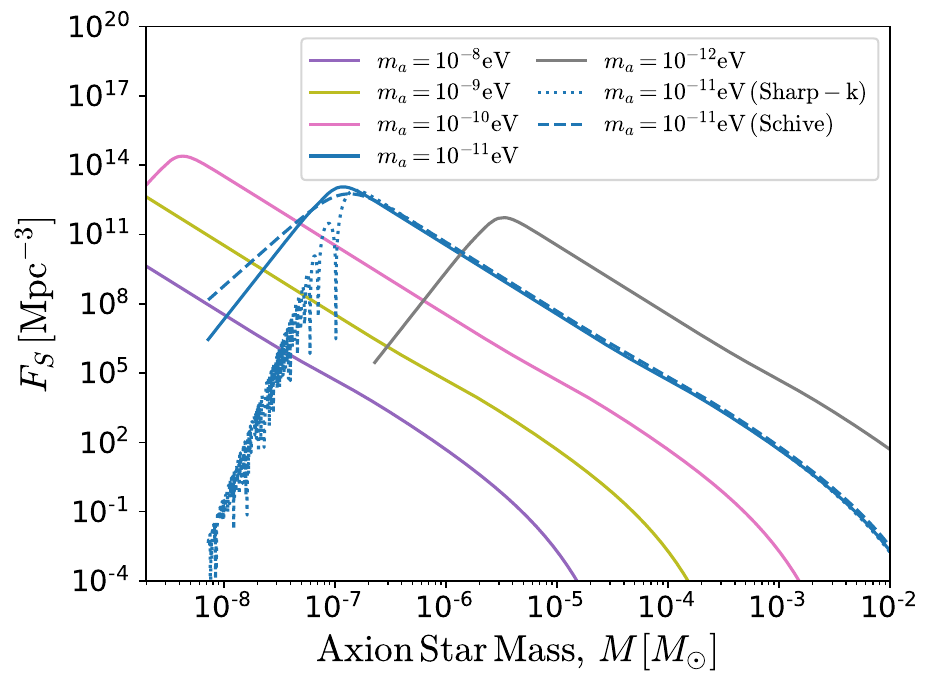}
\caption{The axion star mass function, $F_S={\rm d}n_S/{\rm d}\ln M_S$ defined by Eq.~\eqref{eqn:star-mf}, using the Sheth-Tormen halo mass function at $z=10$, assuming the core-halo mass relation, Eq.~\eqref{eqn:core-halo}. \emph{Left:} in the absence of the Jeans scale, the mass function has a scaling symmetry with all axion star masses proportional to $1/m_a$. We show various values of the exponent $\alpha$ in the core-halo relation. \emph{Right:} including the Jeans scale induces a cut-off and breaks the exact scaling symmetry with $m_a$. We show different models for the cut-off, and different axion masses, holding $\alpha=1/3$ fixed. We use the smooth-$k$ filter as our fiducial window function (solid lines). For $m_a=10^{-11}{\rm eV}$, we also show the axion star mass functions assuming the sharp-$k$ window function (dotted) and that computed from the halo mass function fit found by~\cite{Schive:2015kza} (dashed).
}
\label{fig:star_mf}
\end{center}
\end{figure*}

Simulations of DM structure formation with nonrelativistic scalar fields observe a scaling relation between the DM halo mass, $M_{\rm h}$ and the mass of the central soliton core, $M_c$, known as a ``core-halo mass relation''~\cite{Schive:2014hza}:
\begin{align}
M_{c} = \frac{1}{4}\left[\frac{M_{\rm h}}{M_{\rm min}(z)}\right]^{\alpha}M_{\rm min}(z),
\label{eqn:core-halo}
\end{align}
where $z$ is cosmological redshift, $\alpha$ is a power law exponent, and
\begin{align}
M_{\rm min}(z) = & 1.4\times10^{-6}\left(\frac{m_a}{10^{-13}\text{ eV}}\right)^{-3/2} \nonumber \\
                 &~~~~~\left[\frac{\xi(z)}{\xi(0)}\right]^{1/4}(1+z)^{3/4}M_{\odot}\,,
\label{eqn:Mmin}
\end{align}
is the minimum halo mass at $z$. The function $\xi(z)$ is the virial density contrast~\cite{Bryan:1997dn}. By definition, the total mass of the central soliton/axion star $M_S \approx 4 M_c$~\cite{Schive:2014hza}. The core-halo mass relation can be understood by fixing the soliton radius from the de Broglie wavelength at the virial velocity of the halo with a universal coefficient~\cite{Niemeyer:2019aqm,Eggemeier:2021smj}.

It has been proposed that in Eq.~\eqref{eqn:core-halo} $\alpha=1/3$ is a universal relation in a ``fully relaxed'' DM halo where gravitational energy, kinetic energy, and field gradient energy are in virial equilibrium~\cite{Schive:2014hza,Hui:2016ltb,Bar:2018acw}. The exponent $\alpha=1/3$ is also the attractor solution following multiple mergers~\cite{Du:2016aik}, and may be related to soliton condensation and growth saturation~\cite{Chen:2020cef,Chan:2022bkz}. However, significant scatter in this relationship has also been observed~\cite{Nori:2020jzx,Chan:2021bja,Zagorac:2022xic}, which might be explained by environmental factors, and/or merger history. Recently, it has been argued in Ref.~\cite{Taruya:2022zmt} that the scatter in the halo-concentration relation could also lead to a sizable dispersion in the core-halo mass relation. The effect of scatter on merger rates is discussed in Sec.~\ref{sec:merger_trees}.

We can relate the soliton mass function, $F_S(M_S)=\mathrm{d}n_S/\mathrm{d}\ln M_S$ (where $n_S$ is the soliton number density), to the halo mass function (HMF), $F_{\rm h}(M_{\rm h})=\mathrm{d}n_{\rm h}/\mathrm{d}\ln M_{\rm h}$, by assuming one soliton per halo, $n_S=n_{\rm h}$, i.e.:
\begin{align}
 F_{S}(M_S){\rm d}\ln M_S &= F_{\rm h}(M_{\rm h}){\rm d}\ln M_{\rm h}\, , \\
 \Rightarrow F_S(M_S) &= F_{\rm h}(M_{\rm h}(M_S))\frac{{\rm d}\ln M_{\rm h}}{{\rm d}\ln M_S}\, .
 \label{eqn:star-mf}
\end{align}
Given a core-halo mass relation such as Eq.~\eqref{eqn:core-halo} one can thus compute $F_S$ given $F_{\rm h}$. In the following we compute the halo mass function using the Sheth-Tormen multiplicity function~\cite{Sheth:2001dp}. This analytic approach has been shown to accurately reproduce the soliton mass function found in numerical zoom-in simulations by Ref.~\cite{Eggemeier:2021smj}.

The primary ingredient in the HMF is the halo mass variance, $\sigma^2(M_{\rm h})$, which is computed from the matter power spectrum, $P(k)$, given a window function, $\widetilde{W}(k|M)$. It will turn out that the most phenomenologically interesting halos have low mass, $M\lesssim 1 \,M_\odot$. Accurately and quickly computing the mass variance at low mass requires special care, as described in Appendix~\ref{appendix:calc_details}. 

The power spectrum of scalar field DM displays a low mass cutoff due to the field gradients acting as an effective pressure and inducing a Jeans scale~\cite{Khlopov:1985jw}. The linear power spectrum is strongly suppressed below the Jeans scale at radiation-matter equality~\cite{Hu:2000ke,Marsh:2013ywa}. In $N$-body simulations such as Refs.~\cite{Schive:2015kza,Corasaniti:2016epp}, suppression of the linear power spectrum leads to suppression of halo formation below the half-mode scale of the transfer function relative to CDM. The half-mode scale is redshift independent, since it is determined by initial conditions. On the other hand, the absolute minimum halo mass, Eq.~\eqref{eqn:Mmin}, is determined by the redshift-dependent Jeans scale. The minimum halo mass is in general some orders of magnitude smaller than the half-mode mass.

We compute the power spectrum as described in Appendix~\ref{appendix:calc_details}, which is appropriate for an axionlike field given slow roll-initial conditions, a time-independent particle mass, adiabatic initial perturbations, and initial field value $\phi/f_a\lesssim 1$ (where $f_a$ is the axion decay constant in the interaction Lagrangian given below). Generalization of our results to other models requires the appropriate input $P(k)$, which may differ in shape (see, e.g., Refs.~\cite{Ellis:2022grh,Leong:2018opi,Gorghetto:2022sue}). In models with a small-scale (high-$k$) cut-off to $P(k)$, the HMF depends on the form of the window function used to compute $\sigma$~\cite{Schneider:2013ria}. We investigated both the sharp-$k$ space window function~\cite{Bond:1990iw,Benson:2012su,Schneider:2013ria}, and the smooth-$k$ space window function~\cite{Leo:2018odn,Bohr:2021bdm}, which reproduce results of numerical simulations after so-called ``spurious structure'' is removed. In what follows, we use the smooth-$k$ space window function, since it gives a HMF in better agreement with numerical simulations when $P(k)$ has a sharp cut-off at small scales. 

We show the soliton mass function in Fig.~\ref{fig:star_mf} for various parameters. In the absence of a small-scale cutoff the soliton mass function has a universal form, with the soliton mass scaling inversely with axion mass. In the presence of a cutoff, the soliton mass function does not possess an exact scaling symmetry with $m_a$. We show how our results depend on the slope of the core-halo mass relation, and find that steeper slopes (larger $\alpha$) lead to a higher number density of solitons, with the lowest number density for $\alpha=1/3$. This is due to the mass function possessing a higher number density of low mass halos.  

\begin{figure*}[t]
 \center
  \includegraphics[width=1.7\columnwidth]{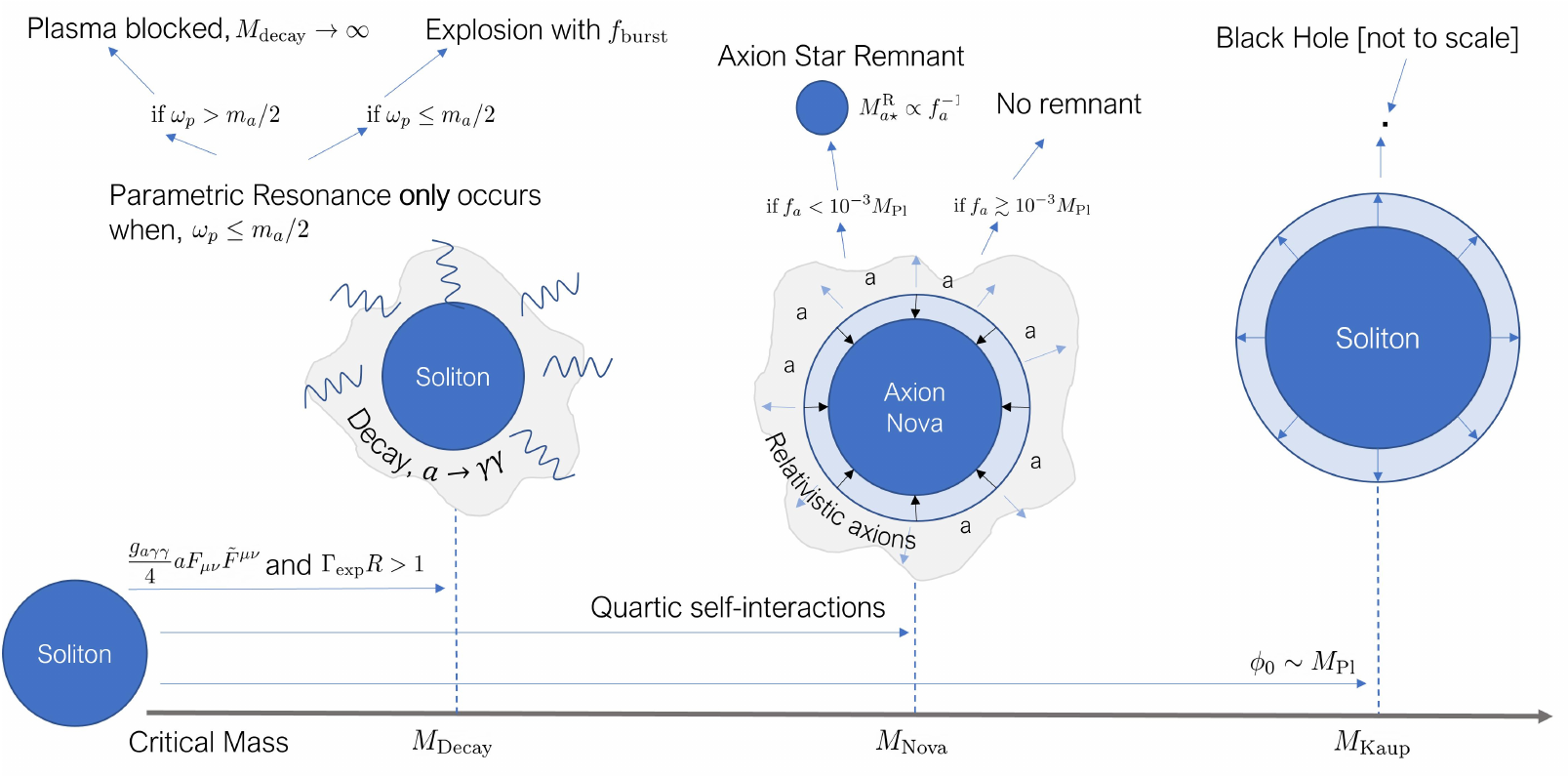}
  \caption{Schematic of critical masses for \textit{three} types of soliton instabilities with the smallest to largest critical masses corresponding to the Decay, Nova and Kaup instability. See Sec.~\ref{sec:critical-soliton} for more detail.}
  \label{fig:decay_channels}
\end{figure*}

\section{Soliton Instability}\label{sec:critical-soliton}

\subsection{Critical Solitons and Critical Halos}
There are several instabilities  shown in Fig.~\ref{fig:decay_channels} that arise when solitons reach a critical mass. Correctly describing the instability requires specifying the interaction Lagrangian, and considering relativistic and environmental effects. All solitons are unstable to black hole formation above the Kaup mass~\cite{Kaup:1968zz}:
\begin{equation}
    M_{\rm Kaup}\approx 0.6\times \frac{M_{\rm Pl}^2}{m_a} \approx 8.5 \times 10^{-2} \,\mathrm{M}_{\odot} \left(\frac{10^{-13}\text{ eV}}{m_a}\right)\, .
    \label{eq:Mkaup}
\end{equation}
This instability occurs when the soliton central field value $\phi_0\sim M_\mathrm{pl}$ (with $M_\mathrm{pl}=1/\sqrt{8\pi G}$ the reduced Planck mass). Soliton collapse to a BH during merger may produce distinct GW signals~\cite{Helfer:2018vtq,Widdicombe:2019woy}. However, as we shall see, ordinary structure formation occurs in environments that are too underdense, and such mergers are expected to be very rare. Soliton collapse to BHs may occur, however, during the period of structure formation after inflation~\cite{Eggemeier:2021smj,DeLuca:2021pls}, or with enhanced primordial fluctuations~\cite{Widdicombe:2018oeo}.

Solitons composed of axions, axion stars, are also unstable due to nonlinear interactions beyond $m_a^2\phi^2$ in the Lagrangian. The relevant terms in the interaction Lagrangian are
\begin{equation}
\mathcal{L}_{\rm int} = -m_a^2 f_a^2 [1-\cos(\phi/f_a)]-\frac{g_{a\gamma\gamma}}{4} \phi F_{\mu\nu}\tilde{F}^{\mu\nu}\,,
\label{eqn:interaction_L}
\end{equation}
where $F_{\mu\nu}$ is the photon field strength tensor, and $\tilde{F}^{\mu\nu}$ is its dual.

In the first term in Eq.~\eqref{eqn:interaction_L}, we have introduced the axion decay constant, $f_a$, which controls the strength of the attractive axion quartic self-interactions, $\lambda$. The presence of these interactions triggers an instability of axion stars above a critical mass. The instability is known as an ``axion Nova," and leads to decay of the DM axion star into relativistic axions~\cite{Levkov:2016rkk,Helfer:2016ljl}. The critical mass for an axion Nova is
\begin{equation}
    M_{\rm Nova}\approx 0.1\,M_\odot  \left(\frac{f_a}{10^{14}\text{ GeV}}\right)\left(\frac{10^{-13}\text{ eV}}{m_a}\right)\, ,\label{eqn:nova-mass}
\end{equation}
The Nova leaves a remnant axion star, with the mass of the remnant depending inversely on the decay constant~\cite{Levkov:2016rkk}. At large decay constant, $f_a\gg 10^{14}\text{ GeV}$, there is no remnant and the total mass of the star is dispersed as relativistic axions. Instability to a Nova occurs at lower mass than the Kaup mass for $f_a\lesssim 0.3 M_\mathrm{pl}$, mapping out a ``phase diagram''~\cite{Helfer:2016ljl,Michel:2018nzt}.

In the second term in Eq.~\eqref{eqn:interaction_L}, we have introduced the axion-photon coupling, $g_{a\gamma\gamma}$. This term also leads to instability of axion stars, driven by parametric resonance~\cite{Hertzberg:2018zte,Levkov:2020txo}.~\footnote{We have recently characterized this mechanism in full numerical relativity simulations~\cite{Chung-Jukko:2023cow}.} The critical mass is
\begin{equation}
    M_{\rm Decay}\approx 8.4\times 10^{-5}M_\odot\left(\frac{10^{-11}\,\text{GeV}^{-1}}{g_{a\gamma\gamma}}\right) \left( \frac{10^{-13}\text{ eV}}{m_a}\right)  \,.\label{eqn:decay-mass}
\end{equation}

The typical expectation for the axion photon coupling, $g_{a\gamma\gamma}$, is that it is related to the decay constant appearing in the self-interaction potential, $f_a$, via $g_{a\gamma\gamma}=\alpha_{\rm EM}/2\pi f_a$, where $\alpha_{\rm EM}$ is the fine-structure constant. Taking $f_a<M_\mathrm{pl}$, we then have $M_{\rm Nova}<M_{\rm Decay}<M_{\rm Kaup}$, and so decay of solitons to relativistic axions via an axion Nova will be the dominant decay channel. However, there are many axion models with enhanced couplings $g_{a\gamma\gamma}$ or suppressed potentials relative to the naive estimate such that soliton decay to photons may be the dominant channel (e.g., ``aligned'' models such as Ref.~\cite{Daido:2016tsj,Daido:2018dmu}, and others~\cite{Sokolov:2022fvs}). Reference~\cite{Levkov:2020txo} further note that the initial collapse of an axion Nova can trigger the parametric resonance instability to photon production. 

\begin{figure}[t]
\begin{center}
\includegraphics[width=\columnwidth]{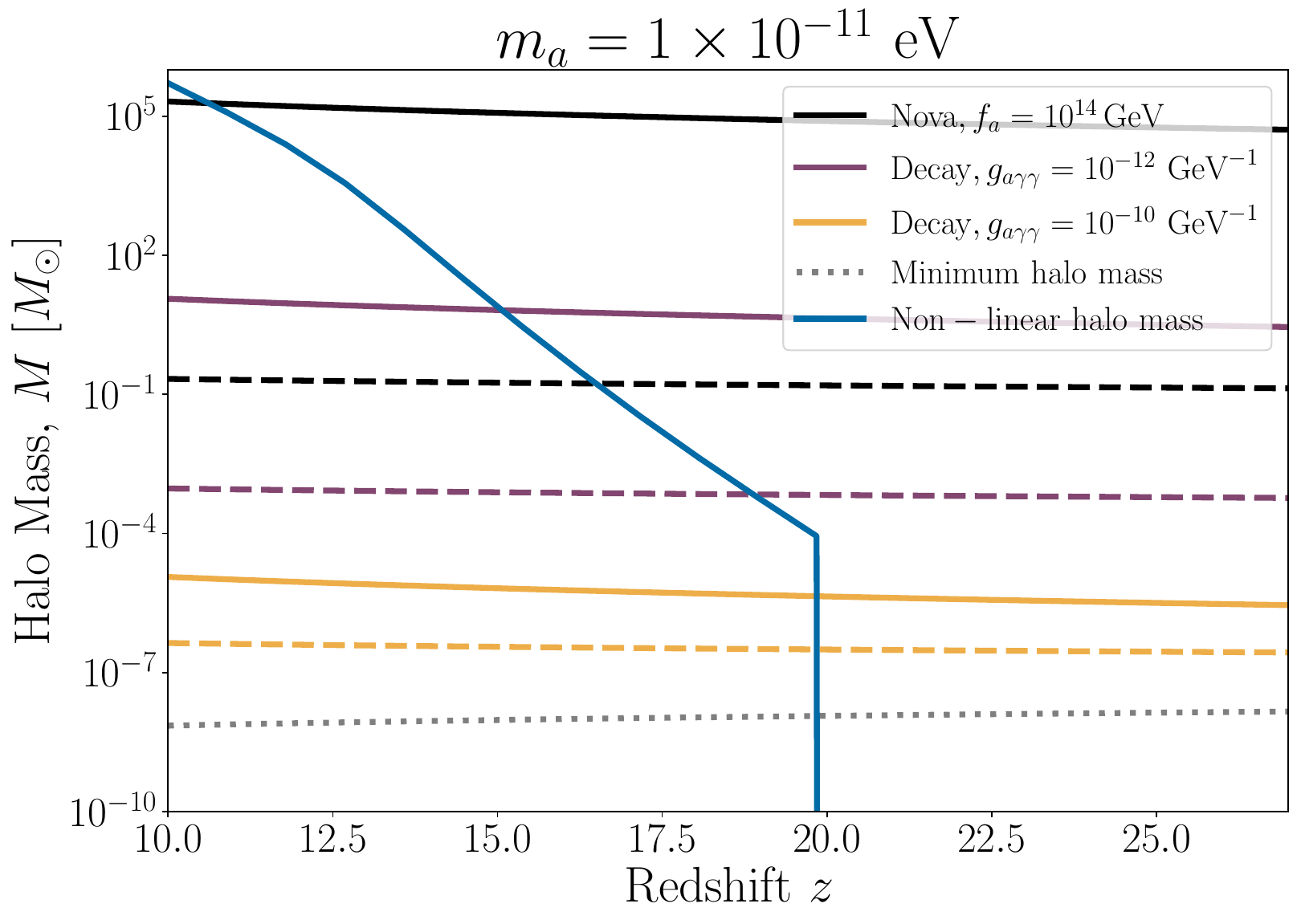}
\caption{Assuming an intrinsic core-halo mass relation, Eq.~\eqref{eqn:core-halo}, with slope parameter $\alpha$, instability of a soliton to Nova [Eq.~(\ref{eqn:nova-mass})] or decay to photons [Eq.~(\ref{eqn:decay-mass})] occurs in halos of a fixed critical mass, as indicated. Solid lines show $\alpha=1/3$, and dashed show $\alpha=3/5$. For reference, we also show the nonlinear mass, $\sigma(M_{\rm nl})=1$, and the minimum halo mass [Eq.~(\ref{eqn:Mmin})]. For the values of the axion-photon coupling and decay constant shown, decay of axion stars to photons occurs before the critical mass for an axion nova can be reached.}
\label{fig:crit_halo}
\end{center}
\end{figure}

Taking any of the instability masses, $M_{\rm Kaup}$, $M_{\rm Nova}$, of $M_{\rm decay}$, rearranging Eq.~\eqref{eqn:core-halo} allows one to find which halos host unstable solitons. Thus, if there is a \emph{universal} core halo mass relation, soliton instability occurs always in halos of a fixed mass with only mild redshift dependence. On the other hand, if there is intrinsic \emph{diversity} in the core-halo mass relation then unstable solitons can occur in a diversity of halos. Furthermore, if $\alpha=1/3$, the critical halo mass is particle mass independent and depends only on the coupling constants.

Assuming a universal core-halo mass relation, it is then useful to compare the critical halo mass hosting unstable solitons to scales in the halo mass function. The mass function has two scales: the minimum halo mass, Eq.~\eqref{eqn:Mmin}, and the nonlinear mass scale, defined by $\sigma(M_{\rm nl})=1$, above which the HMF is exponentially suppressed. Comparing these scales to the critical halo mass gives a rough estimate of when soliton decays are likely to occur in cosmic history, and is shown in Fig.~\ref{fig:crit_halo}. Larger values of $\alpha$ and/or $g_{a\gamma\gamma}$ lead to instability in lower mass, and thus more abundant, halos, and increases the redshift where typical decays occur.  

It was shown in Ref.~\cite{Levkov:2020txo} that if an axion star close to $ M_{\rm Decay}$ grows by slow adiabatic accretion from the background, then it loses energy by efficient photon emission, and returns to a stable condition below $M_{\rm Decay}$. The same is expected to be true for axion emission close to the Nova instability. The growth rate of solitons by accretion is slow, growing as $t^{1/2}$ at low mass~\cite{Levkov:2018kau}, and even slower once virial equilibrium is reached~\cite{Chen:2020cef} (see also \cite{Dmitriev:2023ipv}). Therefore, soliton mass loss by production of photons or axions from dark matter accretion in the host halo is expected to be negligible over a Hubble time. 

There are two methods to obtain rapid decay of solitons:
\begin{itemize}
    \item Plasma blocking;
    \item Major mergers.
\end{itemize}

Plasma blocking effectively sends $M_{\rm Decay}\rightarrow\infty$ as long as $\omega_p(z)>m_a/2$, allowing supercritical solitons with masses larger than Eq.~\eqref{eqn:decay-mass} to form. Once the plasma frequency drops to $\omega_p(z)<m_a/2$ then the decay is kinematically allowed, and supercritical solitons will then decay. We discuss this case in the next subsection. 

Nonadiabatic soliton growth can occur during  major mergers (see, e.g., Ref.~\cite{Hertzberg:2020dbk}). In such a case, a soliton can jump above the critical barrier and will rapidly decay losing an $\mathcal{O}(1)$ fraction of its mass. This process is relevant for decay either to photons or to relativistic axions [i.e. either Eq.~(\ref{eqn:decay-mass}) or~(\ref{eqn:nova-mass})]. We thus expect that the halo major merger rate evaluated at the critical halo mass can be used to estimate the enhanced dark matter decay rate due to soliton mergers. We calculate this in Sec.~\ref{sec:merger_rate}.~\footnote{Reference~\cite{Levkov:2020txo} also computes a maximum relative velocity that solitons can have in order not to Doppler shift the relative frequencies too much and block parametric resonance. This relative velocity is computed for head on soliton collisions. In halo mergers, dynamical friction effects slow down the cores such that mergers occur. As such, we ignore the maximum velocity constraint in the following.} 

\subsection{Soliton Decay to Photons from Plasma Blocking}

Decay of axions to two photons is blocked if the plasma frequency, $\omega_p$, in the environment satisfies $\omega_p>m_a/2$. The low mass DM halos of interest in the present work are below the baryon Jeans scale, and do not possess any cold gas. Therefore, the relevant plasma frequency is given by that of the intergalactic medium (IGM), and is determined by the evolution of the free electron density of the Universe, $n_e(z)$, which is well understood~\cite{Planck:2016mks}. The Universe becomes transparent to photons at recombination, $z\approx 1100$ when $\omega_p\approx 10^{-9}\text{ eV}$. The plasma frequency decreases as the Universe grows and cools, reaching a minimum of $\omega_p \approx 10^{-14}\text{ eV}$ when the first stars begin to reionize the IGM, which occurs between $z=6$ and $z=10$. Thus, post recombination plasma blocking can be neglected for $m_a\gtrsim 10^{-9}\text{ eV}$.~\footnote{Plasma effects on axion photon conversion in neutron star magnetospheres are discussed in e.g Refs.~\cite{Hook:2018iia,Witte:2021arp}. Pre-recombination axion-photon conversion is discussed in Ref.~\cite{Mirizzi:2009iz}.}

When the plasma frequency achieves $\omega_p(z_{\rm crit})=m_a/2$, decay of all super-critical solitons occurs at once, in a burst. The energy released can be expressed as a fraction of the dark matter density at $z_{\rm crit}$ by integrating the soliton mass function: 
\begin{align}
f_{\rm burst}=\frac{1}{\rho_{\rm DM}}\int_{M_{\rm Decay}}^\infty (M_S-M_{\rm Decay})F_S(M_S,z_{\rm crit}){\rm d}\ln M_S\, .
\label{eqn:f_crit_1}
\end{align}
With Eq.~\eqref{eqn:f_crit_1} one can then calculate the fractional dark matter energy density that is converted into photons once $\omega_p(z_{\rm crit})=m_a/2$. However, it is important to highlight that once the energy is injected and transformed into heat and leads to reionization $\omega_p(z_{\rm crit})$ will in turn increase and possibly eventually drop again below $\omega_p(z_{\rm crit-2})=m_a/2$ at a later time, $z_{\rm crit-2} < z_{\rm crit}$. Calculating the subsequent burst(s) energy requires solving for the free electron fraction and baryon temperature: this is addressed in the companion paper Ref.~\cite{Escudero:2023vgv}.

\section{Soliton Mergers and Dark Matter Decay}\label{sec:merger_rate}

\subsection{Halo Formation Rate from Extended Press-Schechter formalism}\label{sec:eps_merger}

To compute the decay rate of axions due to soliton mergers, we need to compute the formation rate of critical axion stars, which is related to the formation rate of corresponding DM halos by the core-halo mass relation. The formation rate of DM halos have been studied extensively in the last decades using analytic formalism and/or numerical simulations. We refer the readers to Refs.~\cite{Lacey:1993iv,Lacey:1994su,1994PASJ...46..427S,Mitra:2011rt} for more detailed discussions. For completeness, we briefly summarize the procedure we have taken below.

The naive halo formation rate comes from the time derivative (redshift derivative) of the mass function. This does not give the correct merger rate, because halos at a specific mass are both formed and destroyed by mergers at the same time. Thus, we can express the redshift derivative of the mass function as~\cite{1994PASJ...46..427S,Mitra:2011rt}:
\begin{equation}
\frac{\mathrm{d}^2 n_{\rm h}(M,z)}{\mathrm{d}M dz}=\frac{\mathrm{d}^2 n_{\rm form}(M,z)}{\mathrm{d}M \mathrm{d}z}-\frac{\mathrm{d}^2n_{\rm dest}(M,z)}{\mathrm{d}M \mathrm{d}z}.
\label{eq:hmf_dt}
\end{equation}
The formation rate of halos per unit mass and per unit volume is given by
\begin{equation}
\!\frac{\mathrm{d}^2 n_{\rm form}(M,z)}{\mathrm{d}M \mathrm{d}z}\!=\!\int_0^{M}\!\frac{M'}{M} \frac{\mathrm{d}n_{\rm h}(M', z)}{\mathrm{d}M'} \frac{\mathrm{d}^2 f_{2\rightarrow 1}(M, M', z)}{\mathrm{d} M \mathrm{d}z} \mathrm{d}M',
\label{eq:formation_rate0}
\end{equation}
with $\mathrm{d}^2 f_{2\rightarrow 1}(M, M', z)/\mathrm{d} M/\mathrm{d}z$ the fraction of mass in a halo of mass $M'$ that merges into halos of mass $M$ at a later time per unit mass and per unit redshift.
Here the subscripts ``1" and ``2" denote halos corresponding to the first ($M$) and second ($M'$) arguments, respectively. The arrow indicates the mass flow, e.g., $2\rightarrow 1$ represents the mass in halo $2$ merges into halo $1$, while $1\leftarrow 2$ represents halo $1$ obtains mass from halo $2$.
On the other hand, the destruction rate of halos per unit mass and per unit volume is given by
\begin{equation}
\frac{\mathrm{d}^2 n_{\rm dest}(M,z)}{\mathrm{d}M \mathrm{d}z}=\int_M^{+\infty} \frac{\mathrm{d}n_{\rm h}(M, z)}{\mathrm{d}M} \frac{\mathrm{d}^2 f_{2\rightarrow 1}(M', M, z)}{\mathrm{d} M' \mathrm{d}z} \mathrm{d}M'.
\label{eq:destruction_rate0}
\end{equation}
Note that in the above equation we do the integration with respect to the descendant halo mass.
One can either compute the halo formation rate directly from Eq.~(\ref{eq:formation_rate0}) or first compute the halo destruction rate from Eq.~(\ref{eq:destruction_rate0}) and then convert it to formation rate using Eq.~(\ref{eq:hmf_dt}). In this work, we choose the former approach.

The mass function and formation rate can be estimated using extended Press-Schechter (EPS) theory~\cite{Bond:1990iw,Lacey:1993iv} as follows. Consider a random overdensity field $\delta(\mathbf{x})\equiv\delta\rho_\mathrm{m}/\rho_\mathrm{m}$ linearly extrapolated to the present time whose power spectrum is $P(k)$. One can smooth out overdensities below the scale $R$ by convolving $\delta(\mathbf{x})$ with a window function $W(\mathbf{x}|R)$:
\begin{equation}
\delta^S(\mathbf{x}) = \int \delta(\mathbf{x}) W(\mathbf{x}+\mathbf{x}'|R) \mathrm{d}^3\mathbf{x}'.
\label{eq:delta_S}
\end{equation}
The variance of $\delta^S(\mathbf{x})$ is given by
\begin{equation}
S(R)\equiv\sigma^2(R) = \int \frac{k^2}{2\pi^2}P(k) \widetilde{W}(k|R)^2 \mathrm{d}k\, ,
\label{eq:sigma_2}
\end{equation}
where $\widetilde{W}(k|R)$ is the Fourier transform of $W(\mathbf{x}|R)$. 

For $R\rightarrow +\infty$, $S=0$ corresponding to $\delta^S(\mathbf{x})=0$ everywhere. As the smoothing scale decreases, more and more perturbations on small scales are included, so $S$ increases. The trajectory of $\delta^S(\mathbf{x})$ at a spatial position is a random walk if we treat $S$ as the time variable. As $S$ increases, $\delta^S(\mathbf{x})$ will eventually pass a certain threshold, $\delta_\mathrm{c}$, which is called the critical overdensity for collapse---the mass element at $\mathbf{x}$ is then considered to be included in a collapsed halo with a mass $M$ that corresponds to the smoothing scale $R$. The relation between $M$ and $R$ depends on the choice of window function (see Appendix~\ref{appendix:calc_details} for details). By computing the probability that $\delta^S(\mathbf{x})$ makes its first upcrossing of $\delta_\mathrm{c}$ at $S(M)$~\footnote{Since $S$ is a monotonic function of $R$ and $R$ is a monotonic function of $M$, here we write $S$ as a function of $M$.}, one can estimate the fraction of mass in the Universe that is contained in halos with a mass $M$. Assuming the spherical collapse model~\cite{Gunn:1972}, the critical overdensity for collapse of CDM (valid for axion DM above the Jeans scale) is mass independent and only a function of redshift. The probability that $\delta^S(\mathbf{x})$ first crosses $\delta_\mathrm{c}$ upward at $S(M)$ per unit $S$ is given by
\begin{equation}
\frac{\mathrm{d}f(S,z)}{dS} = \frac{\delta_\mathrm{c}(z)}{\sqrt{2\pi S}} \frac{1}{S} \exp\left[-\frac{\delta_\mathrm{c}^2(z)}{2 S}\right].
\label{eq:f_PS_CDM}
\end{equation}
A fitting function for $\delta_\mathrm{c}$ in $\Lambda$CDM is given by~\cite{Kitayama:1996ne}
\begin{equation}
\delta_\mathrm{c}(z) = \frac{3(12\pi)^{2/3}}{20}\frac{D(0)}{D(z)}\left[1+0.0123\log_{10}\Omega_\mathrm{m}(z)\right].
\label{eq:delta_c}
\end{equation}
Here $\Omega_\mathrm{m}(z)$ is the fractional matter density at $z$, i.e. $\Omega_\mathrm{m}(z) = \Omega_\mathrm{m,0}(1+z)^3/[\Omega_\Lambda+\Omega_\mathrm{m,0}(1+z)^3]$ and $D(z)$ is the linear growth factor of matter perturbations~\cite{Percival:2005vm}
\begin{equation}
D(z)=\frac{1}{1+z}\,{}_2F_1\left(\frac{1}{3},1,\frac{11}{6},-\frac{\Omega_\Lambda}{\Omega_\mathrm{m,0}(1+z)^3}\right).
\label{eq:D_growth}
\end{equation}
Having $\mathrm{d}f(S,z)/\mathrm{d} S$, the halo mass function can be computed as
\begin{equation}
\frac{\mathrm{d}n_h(M,z)}{\mathrm{d}M} = \frac{\rho_m}{M}\frac{\mathrm{d}f(S,z)}{\mathrm{d} S} \left|\frac{\mathrm{d} S}{\mathrm{d} M}\right|.
\label{eq:HMF_PS}
\end{equation}
Here $\rho_m$ is the mean comoving matter density in the Universe. 

While Eqs.~(\ref{eq:f_PS_CDM}) and (\ref{eq:HMF_PS}) agree reasonably well with $N$-body cosmological simulations, they overpredict the abundance of large halos and underpredicts that of small halos. A more accurate model is to consider ellipsoidal collapse, which gives rise to the Sheth-Tormen fitting function~\cite{Sheth:1999mn,Sheth:1999su} that we have used in Sec.~\ref{sec:mass_func}:
\begin{equation}
\frac{\mathrm{d}f^{\rm ST}(S,z)}{\mathrm{d}S} = A\sqrt{\frac{1}{2\pi}}\sqrt{q}\nu\left[1+(\sqrt{q}\nu)^{-2p}\right]\exp\left(-\frac{q\nu^2}{2}\right)\frac{1}{S},
\label{eq:f_ST}
\end{equation}
where $\nu\equiv \frac{\delta_c(z)}{\sqrt{S}}$, $A=0.3222$, $p=0.3$, and $q=0.707$.

Now let us consider a trajectory starting from $(\delta_c(z),S)$ and then first upcrossing the critical overdensity for collapse at an earlier redshift $z'$ and a larger variance $S'$. This corresponds to a halo $M$ collapsing at $z$ and having a progenitor halo $M'$ at $z'$. Replacing $\delta_c$ and $S$ in Eq.~\eqref{eq:f_PS_CDM} by $\delta_c(z')-\delta_c(z)$ and $S'-S$, we obtain the probability for such an event:
\begin{align}
&\frac{\mathrm{d}f_{1\leftarrow2}(S(M),S'(M'))}{\mathrm{d}S'}= \nonumber \\
&\frac{\delta_c(z')-\delta_c(z)}{\sqrt{2\pi (S'-S)}}\exp\left[-\frac{(\delta_c(z')-\delta_c(z))^2}{2(S'-S) }\right]\frac{1}{S'-S}.
\label{eq:f_1_2_CDM}
\end{align}
Equation~\eqref{eq:f_1_2_CDM} is related to the progenitor mass function of halos of mass $M$, and as we will see, can be used to compute the merger rate, which is found from the reverse conditional probability $df_{2\rightarrow 1}(S, S')/dS$. First, we compute the backward rate, taking a small redshift step $\Delta z$:
\begin{align}
&\frac{\mathrm{d}^2 f_{1\leftarrow2}(S,S')}{\mathrm{d}S' \mathrm{d}z}=\frac{\delta_c(z+\Delta z)-\delta_c(z)}{\Delta z}\times\nonumber \\
&\frac{1}{\sqrt{2\pi (S'-S)}}\exp\left[-\frac{(\delta_c(z+\Delta z)-\delta_c(z))^2}{2(S'-S) }\right]\frac{1}{S'-S}.
\label{eq:f_1_2_CDM_rate}
\end{align}
When $\Delta z\rightarrow 0$, the exponential term in Eq.~(\ref{eq:f_1_2_CDM_rate}) approaches $1$ and thus can be removed from the equation, which is commonly done in the literature (e.g., Ref.~\cite{Lacey:1993iv}), and the term involving $\delta_c$ is simply the derivative with respect to $z$. However, we keep the exponential term and take a small but finite time step size when computing the halo formation rate to avoid divergent results when $S'\rightarrow S$ upon taking integrals over $S'$ (which we do shortly). This is because the quantities are probability distributions, and the order of limits and integrals can be important.

As in the case of the halo mass function a more accurate formula for the rate is obtained by accounting for departures from spherical collapse using an empirical modification calibrated to $N$-body simulations~\cite{Parkinson:2007yh}:
\begin{equation}
\frac{\mathrm{d}^2 f_{1\leftarrow2}^{\rm Nbody}(S,S')}{\mathrm{d}S' \mathrm{d}z} = \frac{\mathrm{d}^2 f_{1\leftarrow2}(S,S')}{\mathrm{d}S'\mathrm{d}z} G(S, S'),
\label{eq:f_Parkinson}
\end{equation}
where $G$ is
\begin{equation}
G(S,S')=G_0\left(\frac{S'}{S}\right)^{\gamma_1/2}\left(\frac{\delta_c(z)^2}{S}\right)^{\gamma_2/2},
\label{eq:branching_mod}
\end{equation}
with $G_0=0.57$, $\gamma_1=0.38$ and $\gamma_2=-0.01$.

We can now find the probability per unit redshift that a halo of mass $M'$ merges into a halo of mass $M$ at a later time~\cite{Lacey:1993iv}:
\begin{equation}
\frac{\mathrm{d}^2 f_{2\rightarrow1}^{\rm Nbody}(S,S')}{\mathrm{d}S \mathrm{d}z}=\frac{\mathrm{d}^2 f_{1\leftarrow2}^{\rm Nbody}(S,S')}{\mathrm{d}S' \mathrm{d}z}\frac{\mathrm{d} f^{\rm ST}(S,z)/\mathrm{d}S}{\mathrm{d} f^{\rm ST}(S',z')/\mathrm{d} S'}.
\label{eq:f_merging}
\end{equation}
This reversal of conditional probabilities follows from conservation of mass, i.e., the mass that halo $S$ gets from halo $S'$ equals to the mass that halo $S'$ merges into halo $S$.

\begin{figure}
\begin{center}
\includegraphics[width=\columnwidth]{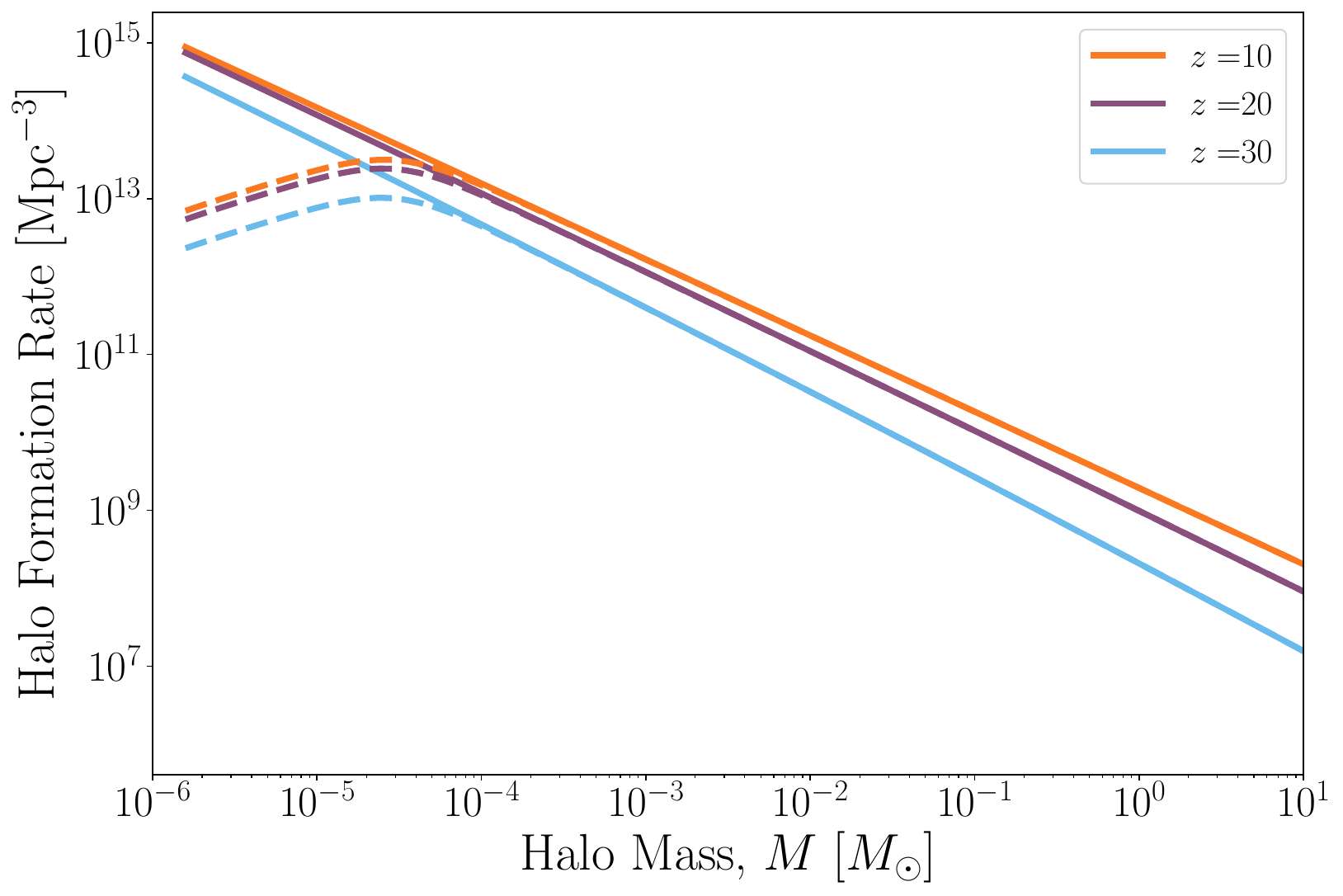}
\caption{Halo formation rate, per logarithmic mass bin, per redshift, with mass resolution $M_{\mathrm{res}}=10^{-3} M$. Solid lines show $m_a=10^{-9}\text{ eV}$, while dashed lines show $m_a=10^{-11}\text{ eV}$. The computation uses the $N$-body fit in Eq.~\eqref{eq:f_Parkinson} from Ref.~\cite{Parkinson:2007yh}, which agrees to within a factor of $\mathcal{O}(3)$ with the EPS result for spherical collapse for $z\gtrsim 10$.}
\label{fig:halo_formation_compare}
\end{center}
\end{figure}

Plugging Eqs.~\eqref{eq:HMF_PS} and \eqref{eq:f_merging} into Eq.~\eqref{eq:formation_rate0}, the halo formation rate is then given by:
\begin{eqnarray}
&& \frac{\mathrm{d}^2 n_{\rm form}}{\mathrm{d}M \mathrm{d}z}\nonumber \\
&=&\int_{S}^{+\infty}\frac{M'}{M}\frac{\rho_m}{M'}\frac{\mathrm{d} f^{\rm ST}(S', z')}{\mathrm{d} S'}
\frac{\mathrm{d}^2 f_{2\rightarrow1}^{\rm Nbody}(S,S')}{\mathrm{d}S \mathrm{d}z} \left|\frac{\mathrm{d}S}{\mathrm{d}M}\right| \mathrm{d} S' \nonumber \\
&=&\frac{ \rho_m }{M} \left|\frac{\mathrm{d}S}{\mathrm{d}M}\right| \int_{S}^{+\infty} \frac{\mathrm{d} f^{\rm ST}(S', z')}{\mathrm{d} S'} \frac{\mathrm{d}^2 f_{2\rightarrow1}^{\rm Nbody}(S,S')}{\mathrm{d}S \mathrm{d}z}\mathrm{d} S' \nonumber \\
&=&\frac{\rho_m}{M} \frac{\mathrm{d} f^{\rm ST}(S,z)}{\mathrm{d} S} \left|\frac{\mathrm{d}S}{\mathrm{d}M}\right| \int_{S}^{+\infty} \frac{\mathrm{d}^2 f_{1\leftarrow2}^{\rm Nbody}(S,S')}{\mathrm{d}S' \mathrm{d}z} \mathrm{d}S' \nonumber \\
&=&\frac{\mathrm{d}n_{\rm h}^{\rm ST}(M,z)}{\mathrm{d} M} \int_{S(M-M_{\text{res}})}^{S_{\rm max}} \!\!\!\! \frac{\mathrm{d}^2 f_{1\leftarrow2}^{\rm Nbody}(S,S')}{\mathrm{d}S' \mathrm{d}z} \mathrm{d}S'.
\label{eq:formation_rate}
\end{eqnarray}
Here in the second and last lines we have used the definition of the halo mass function Eq.~\eqref{eq:HMF_PS}. In the last line, we also introduced finite $S_{\rm max}$ in the maximum of the integral, and finite mass resolution $M_{\rm res}$ in the minimum.
From Eq.~\eqref{eq:formation_rate}, we can see that the halo formation rate is proportional to the halo mass function. 

The upper limit in the integral in Eq.~\eqref{eq:formation_rate}, $S_{\rm max}$, should be the largest value of $S$ (which occurs for $M\rightarrow 0$). For CDM, $S_{\rm max}\rightarrow \infty$.
However, the axion Jeans scale (or indeed the free streaming scale of a thermally produced WIMP) provides a natural upper limit and finite $S_{\rm max}$.
To avoid the divergence of the integral, we only account for halos that merged into $M$ with a mass smaller than $M-M_{\text{res}}$, i.e., replacing the lower limit of the integral with $S(M-M_{\rm res})$. Furthermore, as noted earlier, convergence of the integral requires retaining finite $\Delta z$ in Eq.~\eqref{eq:f_1_2_CDM_rate}, which retains the exponential factor, until after the integral has been performed. Retaining explicit resolution factors everywhere is consistent with the computation of these quantities in $N$-body simulations and in merger trees, which we use to calibrate and check the analytic results.

\begin{figure}
    \centering
    \includegraphics[width=\columnwidth]{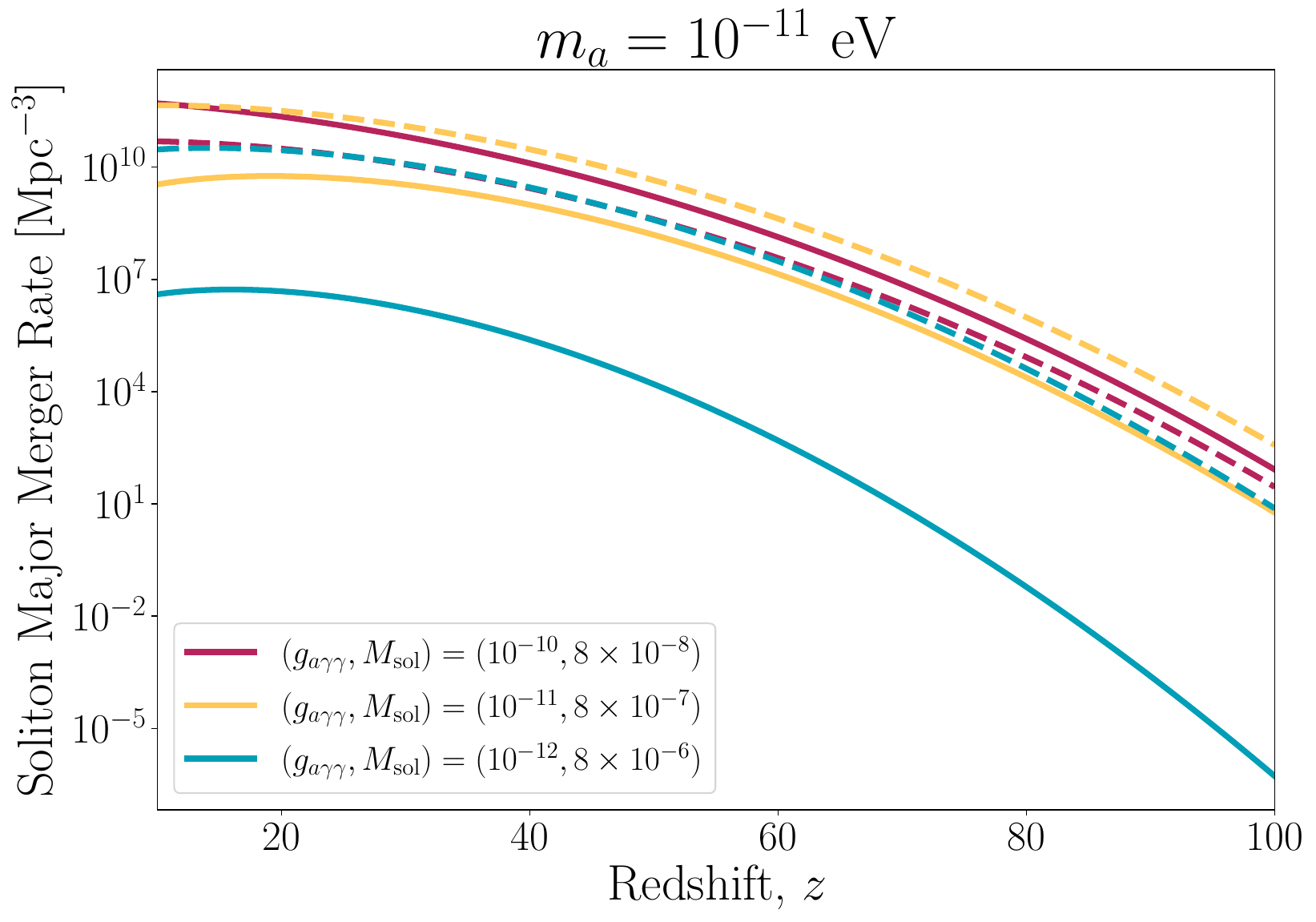}
    \caption{Soliton major merger rate density (per unit redshift, i.e. dimensionless time) for $m_a=10^{-11}\text{ eV}$. The major merger rate is evaluated at a given soliton mass, measured in $M_\odot$, which corresponds to the critical soliton mass for a corresponding value of $g_{a\gamma\gamma}$ (in GeV$^{-1}$). For different applications, the soliton merger rate can be computed at any desired soliton mass regardless of instability/criticality. Solid lines correspond to core-halo mass relation $\alpha=1/3$ and dashed lines to $\alpha=3/5$.}
    \label{fig:soliton_merger}
\end{figure}

Figure~\ref{fig:halo_formation_compare} compares the halo formation rate for two axion masses, $m_a=10^{-9}\text{ eV}$ and $m_a=10^{-11}\text{ eV}$. For $m_a=10^{-11}\text{ eV}$ the effect of the Jeans scale becomes apparent for $M\lesssim 10^{-4}M_\odot$. Comparing to Fig.~\ref{fig:crit_halo}, we notice that although for $m_a=10^{-11}\text{ eV}$ typical fluctuations only become nonlinear for $z<20$, there is still significant halo formation occurring at earlier times from rarer fluctuations (consider the $z=30$ curve in Fig.~\ref{fig:halo_formation_compare}). We further notice that for $m_a=10^{-11}\text{ eV}$ the turn over in halo formation rate at low masses will suppress the formation of critical halos for $g_{a\gamma\gamma}\gtrsim 10^{-10}\text{ GeV}^{-1}$. Bearing in mind these considerations, we now estimate the halo major merger rate, and dark matter decay rate from near-critical soliton mergers.

\subsection{Dark Matter Decay Rate}\label{sec:DM_decay_rate}

As is shown in Ref.~\cite{Schwabe:2016rze}, during a binary merger, the mass of an axion star increases only when the mass ratio of the two axion stars $\mu=M_{c2}/M_{c1} > 3/7$ (assuming $M_{c2} \leq M_{c1}$). Therefore, we will consider only \emph{major mergers} for which the halo mass ratio of two progenitors $\mu_{\rm h} > \mu_{\rm h, min} = (3/7)^{1/\alpha}$, for $M_c\propto M_{\rm h}^{\alpha}$. Then Eq.~\eqref{eq:formation_rate} becomes
\begin{align}
&\frac{\mathrm{d}^2 n_{\rm form}(M,z)}{\mathrm{d}M \mathrm{d}z} = \frac{\mathrm{d}n_{\rm h}(M,z)}{\mathrm{d} M}\times \nonumber\\
&~~~~~~~~~~~\int_{S(M_{\rm upper})}^{S(M_{\rm lower})} \frac{\mathrm{d}^2 f_{1\leftarrow2}^{\rm Nbody}(S,S')}{\mathrm{d}S' \mathrm{d}z} \mathrm{d}S',
\label{eq:formation_rate_major}
\end{align}
where
\begin{align}
M_{\rm upper} &= \min\left\{\frac{1}{1+\mu_{\rm h, max}} M, M_{\rm h,crit}(z)\right\},
\label{eq:mass_upper_upper}\\ 
M_{\rm lower} &= \max\left\{\frac{\mu_{\rm h, min}}{1+\mu_{\rm h, min}} M, M-M_{\rm h,crit}(z)\right\}.
\label{eq:mass_lower_upper}
\end{align}
In the above equations, we have imposed another condition that the mass of two progenitor halos are both smaller than the critical halo mass.

Additionally, since the halo mass after a merger is not necessarily exactly equal to the critical halo mass, e.g., two halos with masses below the critical halo mass merge and form a bigger halo with a mass larger than the critical value, we integrate~\eqref{eq:formation_rate_major} over the interval $(M_{\rm h,crit}, 2M_{\rm h,crit})$ to get the total number of halos that produce axion stars above the critical mass between redshifts $z+\Delta z$ and $z$. 

We thus arrive at a key result: the major merger rate of solitons around a given critical mass is given by
\begin{widetext}
\begin{equation}
\frac{\mathrm{d}n_{\rm merge}}{\mathrm{d}z} = \int_{M_{\rm halo, crit}}^{2 M_{\rm halo, crit}} \frac{\mathrm{d}n_{\rm h}(M,z)}{\mathrm{d} M} \int_{S(M_{\rm upper})}^{S(M_{\rm lower})} \frac{\mathrm{d}^2 f_{1\leftarrow2}^{\rm Nbody}(S,S')}{\mathrm{d}S' \mathrm{d}z} \mathrm{d}S' \mathrm{d}M.
\label{eq:star_formation_rate}
\end{equation}
\end{widetext}
An example of the soliton major merger rate is shown in Fig.~\ref{fig:soliton_merger}. The rate is computed for a given soliton mass $M_{\rm sol}$. The mass is chosen to correspond to the critical mass for decay to photons, Eq.~\eqref{eqn:decay-mass} for various values of $g_{a\gamma\gamma}$. However, the same rate calculation can be used for any soliton mass of interest. We also demonstrate dependence on the core-halo mass relation parameter, $\alpha$. 

\begin{figure*}
\begin{center}
\includegraphics[width=\columnwidth]{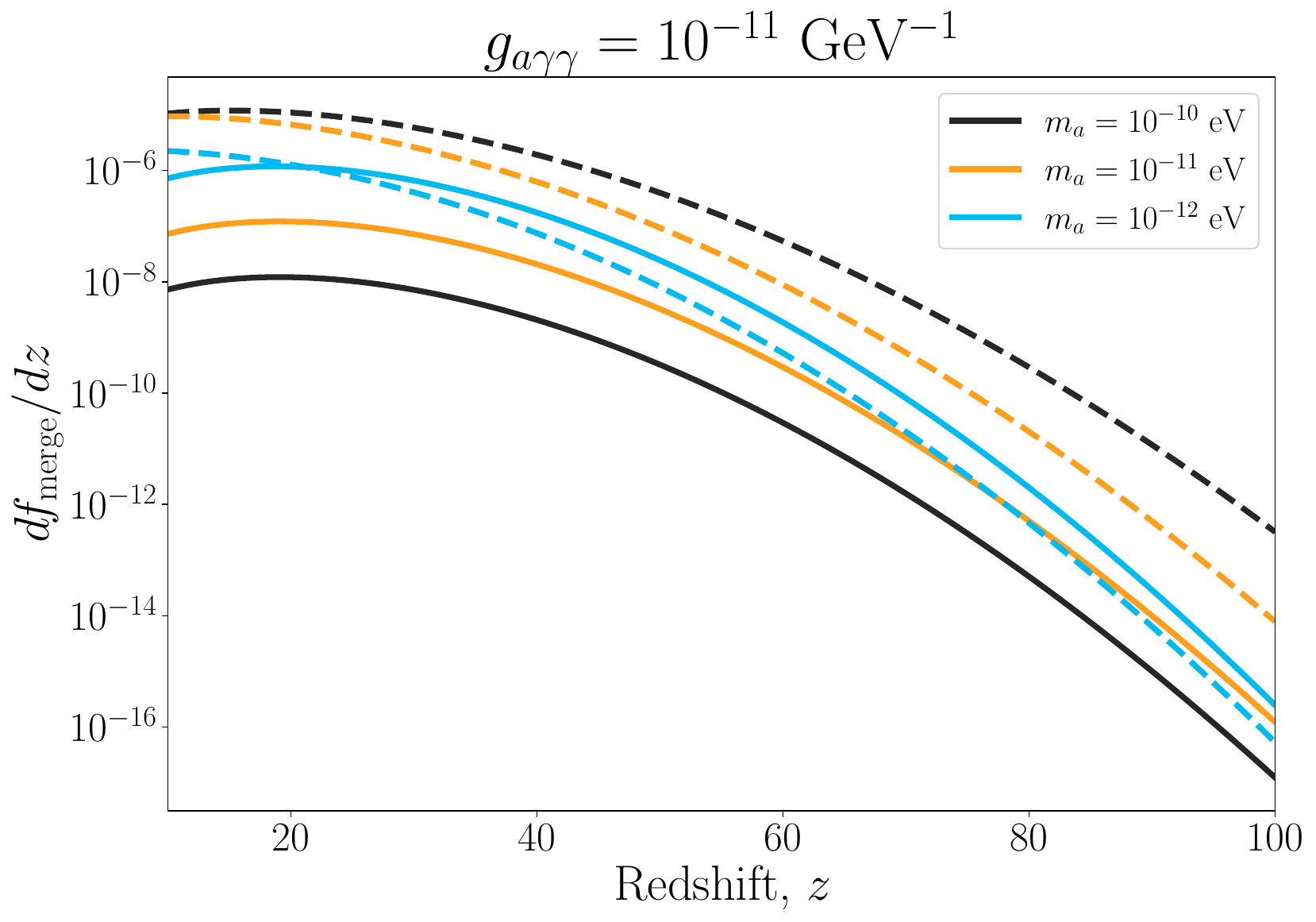}
\includegraphics[width=\columnwidth]{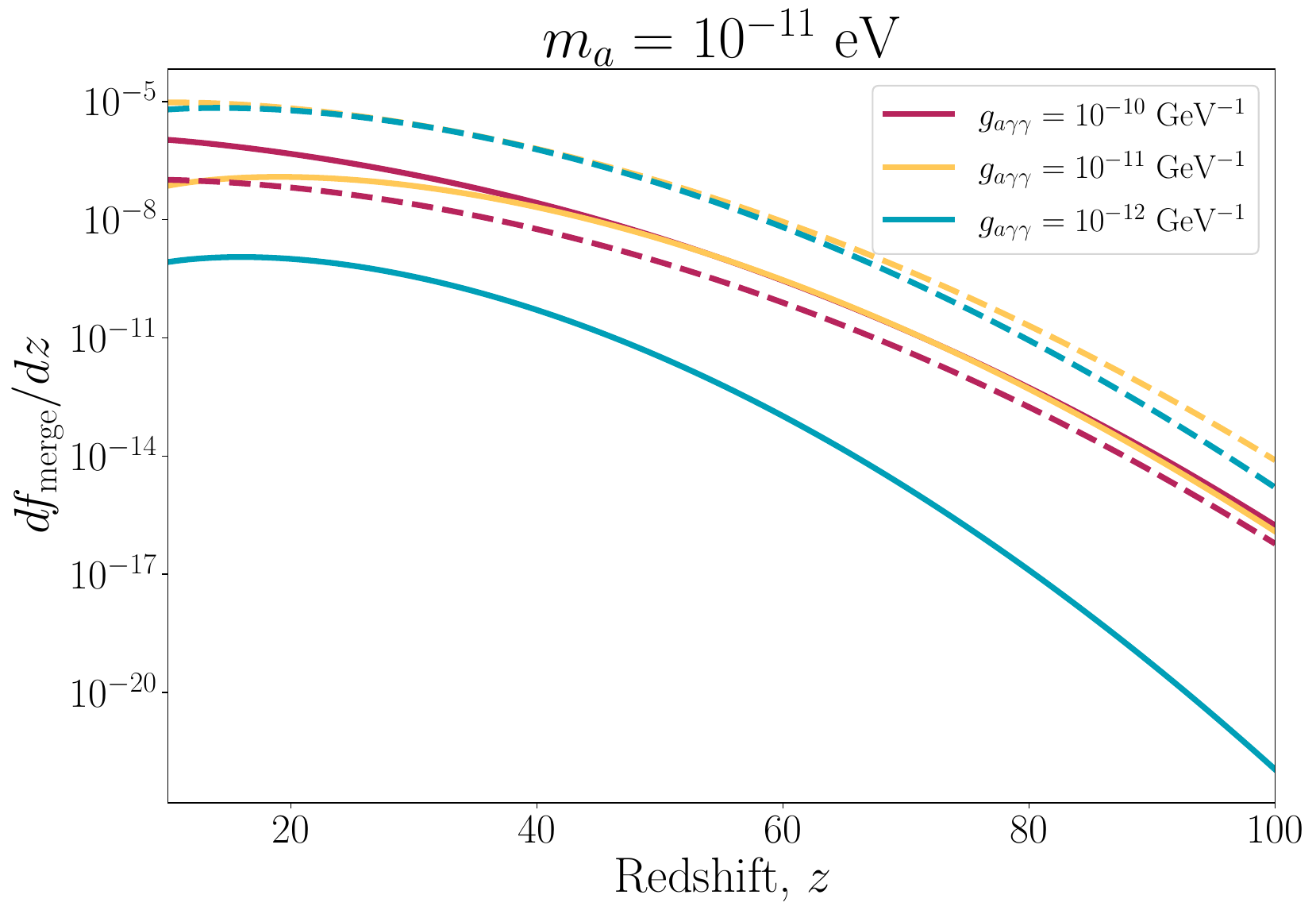}
\caption{Axion dark matter fractional decay rate to photons due to soliton major mergers and parametric resonance. Solid lines assume a core-halo mass relation with slope $\alpha=1/3$, while dashed lines show $\alpha=3/5$. \emph{Left:} dependence on axion mass at fixed coupling. \emph{Right:} dependence on coupling at fixed axion mass.}
\label{fig:df_dz_compare}
\end{center}
\end{figure*}

We can reexpress Eq.~\eqref{eq:star_formation_rate} in terms of the fractional decay rate of dark matter, $\mathrm{d}f_{\rm merge}/\mathrm{d}z$: 
\begin{equation}
    \frac{\mathrm{d}f_{\rm merge}}{\mathrm{d}z} = \frac{M_{\rm S,crit}}{\rho_m} \frac{\mathrm{d}n_{\rm merge}}{\mathrm{d}z} . 
    \label{eqn:df_dz}
\end{equation}
Note that here we have neglected an $\mathcal{O}(1)$ factor and assumed that the axion star completely decays. For $M_{\rm half} < M_{\rm h, crit} < 10^5{M_{\odot}}$ (where $M_{\rm half}$ is the mass corresponding to the half-mode scale in the transfer function relative to CDM), Eq.~\eqref{eqn:df_dz} can be well approximated by
\begin{equation}
\frac{\mathrm{d}f_{\rm merge}}{\mathrm{d}z} = \left. \chi \frac{M_{\rm S,crit}}{\rho_m} \frac{\mathrm{d}n_{\rm h}}{ \mathrm{d}\ln M}\right|_{M=M_{\rm h, crit}},
\label{eq:star_formation_rate_approx}
\end{equation}
where $\chi$ is an $\mathcal{O}(0.1)$ coefficient 
\begin{eqnarray}
\chi = \frac{a(z) + b(z) \left[\log_{10}\left(\frac{M_{\rm h,crit}}{10^{-6}M_{\odot}}\right)\right]}
            {1+ c(z) \left(\frac{M_{\rm h,crit}}{10^{5}M_{\odot}}\right)^{0.39}}.
\label{eq:xi}
\end{eqnarray}
The parameters $a$, $b$ and $c$ are redshift dependent:
\begin{eqnarray}
a &=& a_0 + a_1 z + a_2 z^2, 
\label{eq:fit_par_a} \\
b &=& b_0 + b_1 z + b_2 z^2, 
\label{eq:fit_par_b} \\
c &=& c_0 + c_1 z + c_2 z^2.
\label{eq:fit_par_c}
\end{eqnarray}
The best-fit coefficients for several different values of $\alpha$ are given in Table~\ref{table:fit}. 

\begin{table}[t]
\begin{tabular}{c|c|c|c}
\hline
 ~~~~~~~~  &~~~~$\alpha=1/3~~~~$ &  $~~~~\alpha=2/5~~~~$ &  $~~~~\alpha=3/5~~~~$ \\
\hline
 $a_0$ &   1.07369E-1  &   9.86049E-2  &    8.00409E-2   \\
 $a_1$ &  -7.55082E-5  &  -7.07799E-5  &   -6.22972E-5   \\
 $a_2$ &  -9.39435E-7  &  -9.54236E-7  &   -9.31684E-7   \\
 \hline
 $b_0$ &   6.84693E-3  &   6.28750E-3  &    5.10077E-3   \\
 $b_1$ &  -1.40186E-5  &  -1.36381E-5  &   -1.31868E-5   \\
 $b_2$ &  -1.85082E-7  &  -1.86434E-7  &   -1.68590E-7   \\
 \hline
 $c_0$ &   2.71084E-1  &   2.65829E-1  &   2.57776E-1    \\
 $c_1$ &   2.47512E-3  &   2.67660E-3  &   2.76791E-3    \\
 $c_2$ &   6.89772E-5  &   7.77110E-5  &   1.02358E-4    \\
 \hline
\end{tabular}
\caption{Best-fit coefficients in Eqs.~\eqref{eq:fit_par_a},~\eqref{eq:fit_par_b} and~\eqref{eq:fit_par_c} for different values of slope parameter $\alpha$.}
\label{table:fit}
\end{table}

Figure~\ref{fig:df_dz_compare} shows the DM fractional decay rate, $\mathrm{d}f_{\rm merge}/\mathrm{d}z$, for a variety of model parameters. For fixed axion parameters $m_a$ and $g_{a\gamma\gamma}$ there is a large dependence on the core-halo mass relation, expressed via $\alpha$. For $\alpha=1/3$ we notice almost constant power law scaling of the result with $m_a$ and $g_{a\gamma\gamma}$.

\begin{figure}[t]
\begin{center}
\includegraphics[width=\columnwidth]{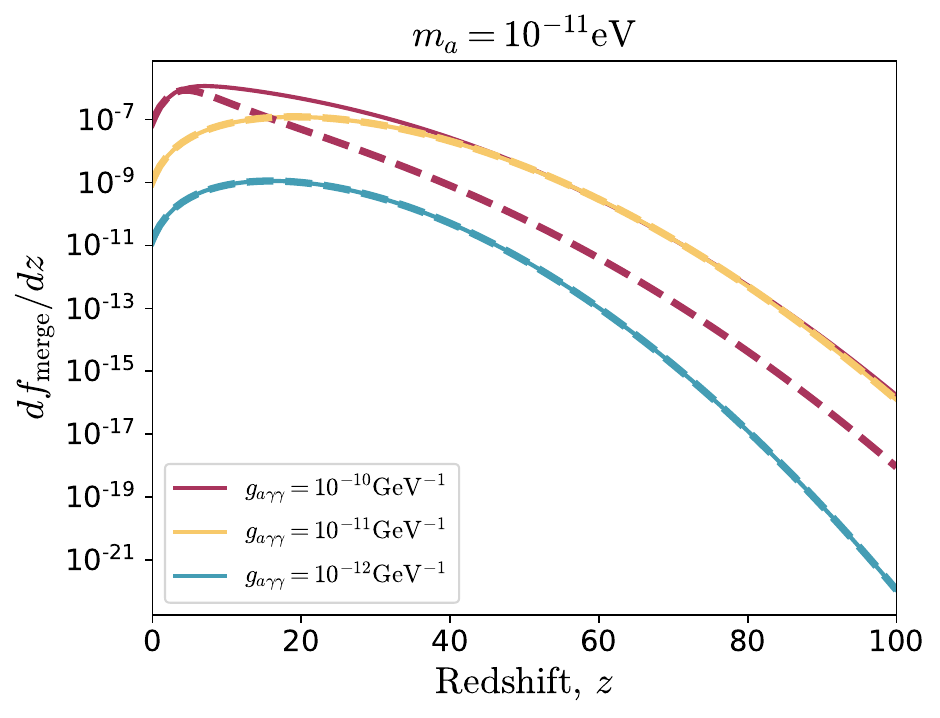}
\caption{Comparison of fractional dark matter decay rate due to soliton mergers comparing the full EPS calculation, Eq.~\eqref{eqn:df_dz} (solid lines), to the fitting formula Eq.~\eqref{eq:star_formation_rate_approx} (bold dashed lines). We fix the axion mass  $m_a=10^{-11}{\text{eV}}$ and core-halo mass relation $\alpha=1/3$. }
\label{fig:df_dz_fit}
\end{center}
\end{figure}

Figure~\ref{fig:df_dz_fit} compares the DM fractional decay rate from the full EPS calculation to the fit given in Eq.~\eqref{eq:star_formation_rate_approx}. The fit is in general very good, except that the fit breaks down at high values of $g_{a\gamma\gamma}\gtrsim 10^{-10}\text{ GeV}^{-1}$ when the critical halo mass drops below $M_{\rm half}$ where the halo mass function shape changes abruptly. Such high values of $g_{a\gamma\gamma}$, however, are robustly excluded by nonobservation of solar axions by the Cern Axion Solar Telescope~\cite{CAST:2017uph}. We conclude that the fit in Eq.~\eqref{eq:star_formation_rate_approx} can be used quite generally.

The decay rate, Eq.~\eqref{eqn:df_dz} can be converted into energy per unit volume per unit time with appropriate factors of the Hubble parameter, $H(z)$, and dark matter density $\Omega_{\rm DM}h^2=0.12$. The energy injection is shown in Fig.~\ref{fig:energy}, compared to a typical energy injection from supernovae. For the supernova model we use the results of Ref.~\cite{Hartwig:2022lon} for core-collapse Pop-III supernovae. We approximate the star formation rate density as a constant between redshifts $10$ and $30$, and assume one core collapse supernova per $100$ Solar masses of star formation. We observe that for the axion parameters considered the energy injection into the intergalactic medium caused by axion star explosions is significantly larger than the energy injection due to supernovae, and with energy injection extending significantly into the dark ages at $z\gg 20$, suggesting that this is a phenomenon with observable consequences.

The python code we used to do the calculations in this subsection is publicly available at \url{https://github.com/Xiaolong-Du/Merger_Rate_of_Axion_Stars}.

\subsection{Merger Trees}\label{sec:merger_trees}

In the previous subsection, we showed the calculations of DM fractional decay rate for a given power-law core-halo mass relation using the EPS formalism. In reality, there can be a large dispersion in the core-halo mass relation~\cite{Chan:2021bja} and a universal core-halo mass relation may not exist~\cite{Zagorac:2022xic}. Thus for fixed axion parameters, $m_a$ and $g_{a\gamma\gamma}$, the critical halo mass also has a scatter making it difficult to compute the DM fractional decay rate using Eqs.~\eqref{eq:star_formation_rate} and~\eqref{eqn:df_dz}. The lower and upper limits of the integral are not determined by a unique core-halo mass relation, e.g., the halo and its progenitors may have different critical values. One way to include possible variation in the core-halo mass relation is to build a large number of ``merger trees'' (Monte Carlo realizations of halo formation based on EPS)~\cite{Lacey:1993iv,Parkinson:2007yh} which record the merger history of halos. Then each halo can be assigned a core mass from a distribution function or following the core mass growth model proposed by one of us in Ref.~\cite{Du:2016aik}. 

Being Monte Carlo models, merger trees also allow for additional physics to be included within individual halos. In our case, we allow that the axion stars are removed from halos after they cross criticality, which circumvents a possible double counting in our calculation of the soliton merger rate from the halo merger rate using EPS. For example, when computing the integral Eq.~\eqref{eq:formation_rate_major}, we consider only the cases with progenitor halo masses smaller than $M_{\rm h,crit}(z)$ [see Eqs.~\eqref{eq:mass_upper_upper} and~\eqref{eq:mass_lower_upper}]. However, $M_{\rm h,crit}(z)$ decreases with increasing redshift (increases with time), so there might be cases where the progenitor halo mass is above the critical value at an earlier time. Those cases should not be included in the integral Eq.~\eqref{eq:formation_rate_major}.

\begin{figure}[t]
\begin{center}
\includegraphics[width=\columnwidth]{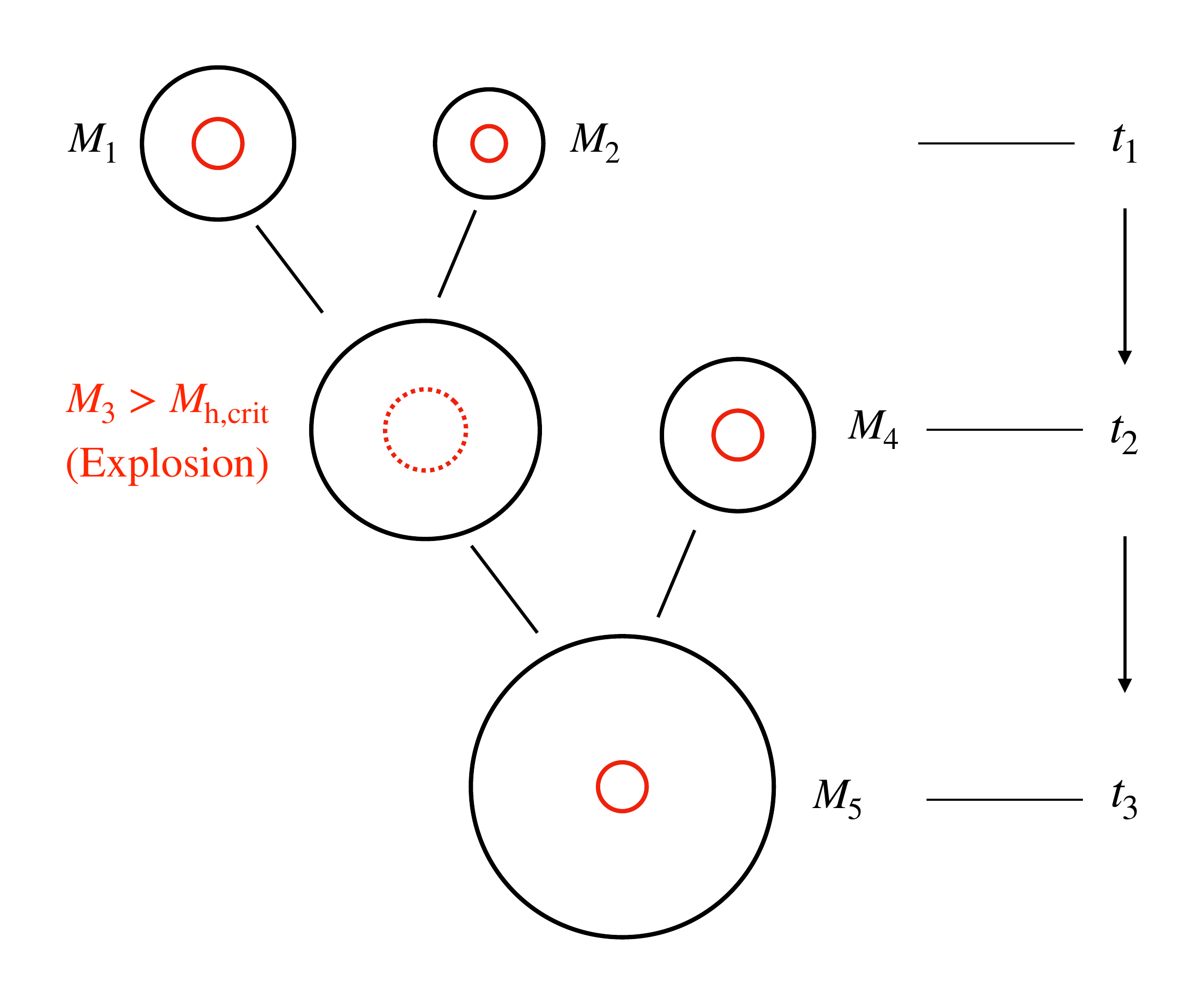}
\caption{A sketch of merger tree. At time $t_1$, two halos (black circles) $M_1$ and $M_2$ are below the critical halo mass. Each halo has an axion star in its center (red circles). At time $t_2$, $M_1$ and $M_2$ merge and form a larger halo $M_3 > M_{\rm h, crit}$. The axion star in the center of $M_3$ (dashed red circle) becomes unstable and explodes. Thus we remove the central axion star from $M_3$, i.e. resetting the core mass to $0$. At a later time $t_3$, halo $M_3$ merges with another halo $M_4$ (below the critical halo mass). If the central axion star in $M_4$ is not disrupted during the merger, the final halo $M_5$ will have a new core (central axion star).}
\label{fig:merger_tree}
\end{center}
\end{figure}

Figure~\ref{fig:merger_tree} shows the schematic diagram of a merger tree. Two halos, $M_1$ and $M_2$, below $M_{\rm h,crit}$ merge and form a larger halo $M_3 = M_1 + M_2 + \Delta M_{\rm acc}$ whose mass is larger than $M_{\rm h,crit}$. Here $\Delta M_{\rm acc}$ is the mass accreted from sub-resolution halos and the mass smoothly accreted from the environment (see Refs.~\cite{Benson:2012su,Du:2016zcv} for more details). After the merger, we assume the central axion star in $M_3$ completely decays and remove it from the halo. In reality, when two halos merge, it will take some time for their central axion stars to merge and form a new axion star, which may lead to delayed axion decays. The timescale for the merging process is discussed in Appendix~\ref{appendix:time_scale}. We leave more detailed study of such a delay effect to further work. We also do not consider the reformation of axion star after the first explosion through gravitational relaxation since this process is slow~\cite{Levkov:2018kau,Chen:2020cef} compared to mergers, but we do allow the halo to accrete a new axion star at a later time through major merger in some of the models we consider below. This essentially changes the core-halo mass relation above the critical halo mass.

\begin{figure}[hbtp]
\begin{center}
\includegraphics[width=\columnwidth]{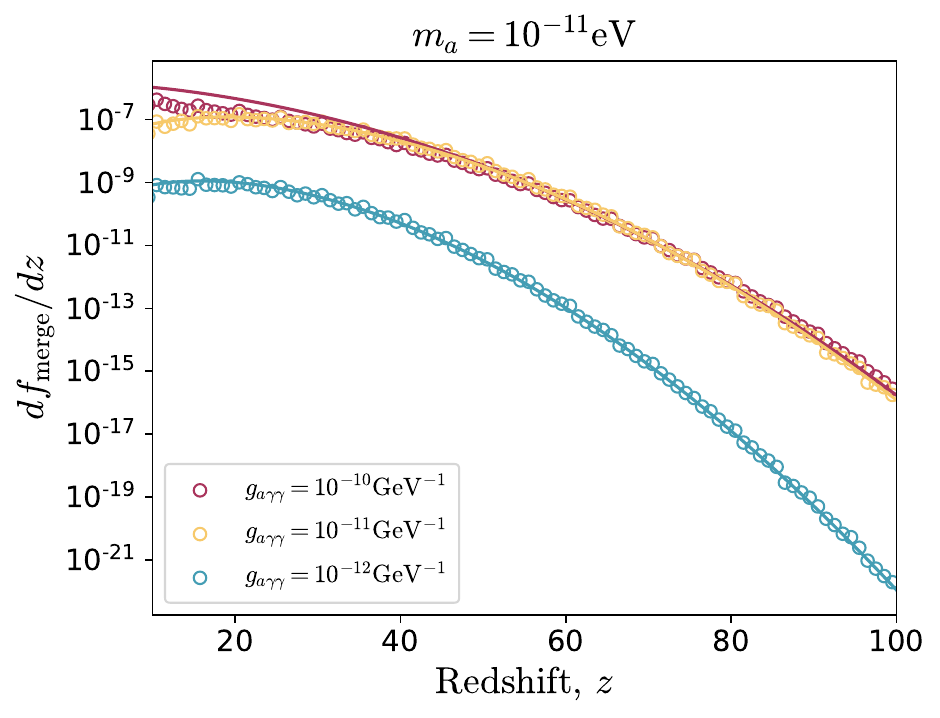}
\caption{Comparison of mean DM fractional decay rate from merger trees (empty circles) with that from EPS formalism (solid lines).}
\label{fig:dfdzfigures}
\end{center}
\end{figure}

We use the semi-analytic code \textsc{Galacticus}~\cite{Benson_2012} to generate realizations of merger trees. More details about the merger tree setup can be found in Appendix~\ref{appendix:merger_tree}. To cross-check with the calculations in the previous subsection, we first look at a fixed core-halo mass relation without any scatter. As mentioned previously, the core-halo mass relation is changed due to the decay of supercritical axion stars. So the core-halo mass relation is not applied to those halos that were ever above the critical halo mass in the past. We make a naive assumption that their core masses remain $0$ since the first explosion and leave the study of other details such as the formation of a new core through dynamical relaxation to future work. The results of this calculation are shown in Fig.~\ref{fig:dfdzfigures}. We see good agreement between the analytic EPS model and the merger tree. This indicates that the calibration of the halo formation and destruction rates was performed correctly, and that there was no double counting of solitons caused by their nonremoval after explosion in the EPS result.

To introduce scatter to the core-halo mass relation, we consider two models: (1) determining the core mass based on the merger histories as in Ref.~\cite{Du:2016aik}; (2) sampling the slope of the core-halo mass relation from a Gaussian distribution such that the core masses have a large scatter as found by Ref.~\cite{Chan:2021bja}. In the first model, the core mass grows as halo mass increases. However, this is not guaranteed in the second model. So in the second model after we draw a core mass for the halo in the merger tree, we checked whether it is smaller than the core mass of the halo's main progenitor. If it is, the halo is assigned the same core mass as its main progenitor. This additional constraint tends to make the slope of the core-halo mass relation steeper as mergers happen. To recover the core mass range found by Ref.~\cite{Chan:2021bja}, we set the mean and standard deviation of the slope parameter as $\alpha_{\rm mean}=0.326$~\footnote{Note that the mean value required for the dynamical merger tree model to reproduce the scatter is smaller than the measured mean value at fixed $z$, $0.515$, reported in Ref.~\cite{Chan:2021bja}.} and $\sigma_{\alpha}=0.1$.

\begin{figure}[t]
\begin{center}
\includegraphics[width=\columnwidth]{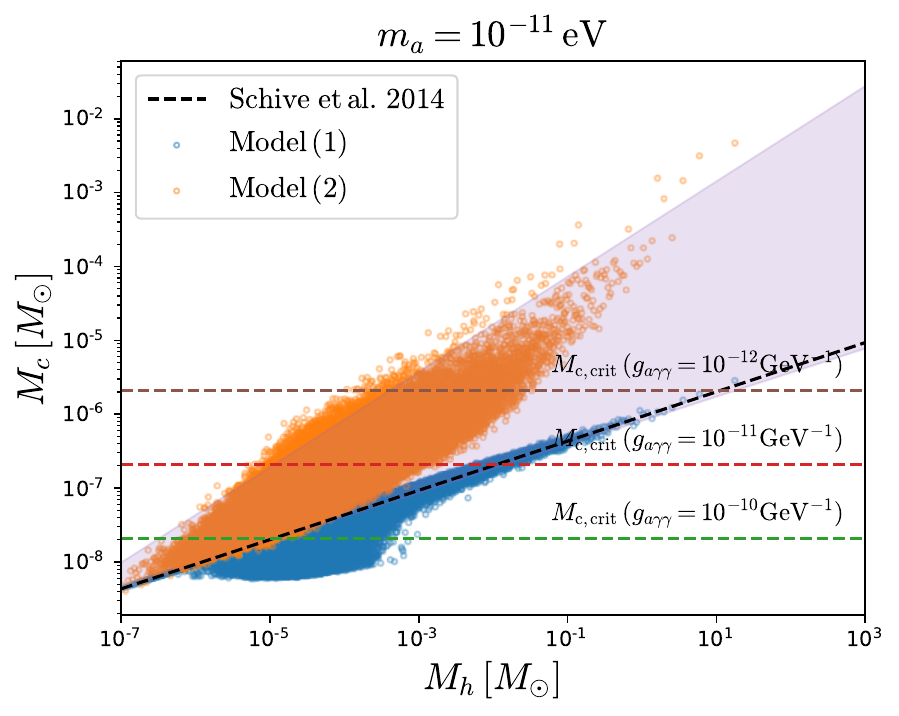}
\caption{Core mass versus halo mass at $z=10$ for $m_a=10^{-11}\,{\rm eV}$ from different models: (1) Du {\it et al.} 2017 ~\cite{Du:2016aik} (blue circles); (2) random sampling (orange circles). The shaded region shows the range of core masses found by~\cite{Chan:2021bja}. The horizontal lines mark the critical core mass $M_{\rm c, crit}=M_{\rm decay}/4$.}
\label{fig:Mc_Mh}
\end{center}
\end{figure}

Figure~\ref{fig:Mc_Mh} shows the core mass versus halo mass from the two models above. In this test, to compare with previous findings on the core-halo mass relation, we have not included the decay of super-critical axion stars. In model (1), the core mass grows only in a major merger event, so for small halos which have experienced only a few mergers the core masses have a large scatter, i.e. a halo growing through major mergers will have a large core mass than the one growing through minor mergers. As more mergers happen, the core-halo mass relation approaches a $1/3$-power law, which is consistent with that found by Ref.~\cite{Schive:2014hza} in cosmological simulations, but the scatter in core masses is much smaller than the other model. As we expected, model (2) reproduces the dispersion in the core-halo mass relation reported in Ref.~\cite{Chan:2021bja} (shaded region).

Then we rerun the above two models and allow the axion star to decay when it becomes supercritical. For model (2), as in the case with a fixed core-halo mass relation, we apply Eq.~\eqref{eqn:core-halo} only to halos that have never been above the critical halo mass. For model (1), we allow the halo to accrete a new core from major merger, but assume there is no adiabatic core growth.

\begin{figure}[t]
\begin{center}
\includegraphics[width=\columnwidth]{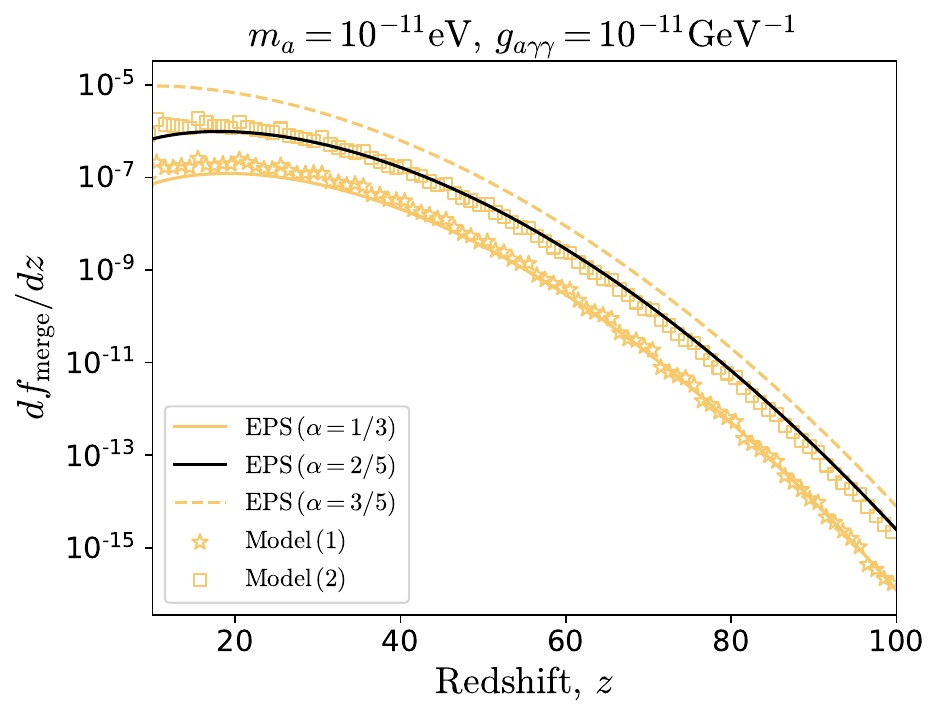}
\caption{DM fractional decay rate from merger trees assuming different core-halo mass relation models comparing with the EPS models assuming a fixed core-halo mass relation with different values of $\alpha$.
}
\label{fig:df_dz_scatter}
\end{center}
\end{figure}

In Fig.~\ref{fig:df_dz_scatter}, we compare the DM fractional decay rate predicted by the models above with those assuming a fixed core-halo mass relation. We found that the results from model (1) agree well with a fixed core-halo mass relation with $\alpha=1/3$, while the results from model (2) lies between two limiting cases with $\alpha=1/3$ and $\alpha=3/5$, corresponding to an effective average value $\alpha\approx2/5$.

\section{Summary and Conclusion}\label{sec:conclusions}

For DM composed of a bosonic field, numerical work in the last decade has shown that a general prediction of such a model is the formation of a soliton in every dark matter halo. Numerical and semi-analytic models can be used to compute the soliton mass function and merger rate. In the present work, we used the extended Press-Schechter formalism to first write down the soliton mass function assuming a core-halo mass relation with the host DM halos. Figure~\ref{fig:star_mf} shows the resulting number density of solitons predicted in a standard cosmology. 

Using this result as a baseline, we then presented a new calculation of the soliton formation rate and merger rate, culminating in the double-differential formation rate, Eq.~\eqref{eq:formation_rate}. Solitons form and grow primarily due to mergers, and the formation rate can be used to calculate the rate of \emph{major mergers} of solitons of any given mass, Eq.~\eqref{eq:star_formation_rate}, and Fig.~\ref{fig:soliton_merger}.

Due to nonlinear interactions, and including relativistic corrections, bosonic DM solitons are unstable above a critical mass. For gravity, this instability leads to BH formation, but is predicted to occur only in very massive DM halos. For quartic self-interactions, and for axionlike interactions with photons, the instability leads to soliton explosions, and decay of DM into relativistic degrees of freedom. Soliton explosions caused by major mergers producing heavy and unstable solitons thus leads to an enhanced DM decay rate. 

We write the decay rate as $\mathrm{d}f/\mathrm{d}z$, where $f$ is the fraction of the total DM density, and $z$ is redshift. Assuming a given core-halo mass relation, our computation of this rate can be approximated from the halo mass function and is given in Eq.~\eqref{eq:star_formation_rate_approx}, which applies for any instability to decay at a mass $M_{\rm decay}$, regardless of the specific mechanism. Taking the axion-like instability to production of radio photons~\cite{Levkov:2020txo}, we used our computation of the major merger rate to compute the DM decay rate for different values of the axion mass, $m_a$, and coupling constant, $g_{a\gamma\gamma}$, as shown in Fig.~\ref{fig:df_dz_compare}. 

The analytical EPS model can only be used to compute the DM decay rate if there exists an exact analytical core-halo mass relation. Given that this is thought to be a statistical phenomenon, with scatter due to halo formation histories and departures from equilibrium, we extended our computation to semi-analytical Monte Carlo methods using the \textsc{Galacticus} code~\cite{Benson_2012}. Firstly, we used our Monte Carlo model to validate the EPS calculation of the merger rate for a fixed core-halo mass relation. Next, we introduced scatter to the core halo mass relation tuned to match results of numerical studies~\cite{Chan:2021bja}, as shown in Fig.~\ref{fig:Mc_Mh}. We found that the results of such a model can be matched by the EPS model using a mean value of the core-halo mass relation slope parameter $\alpha$, as shown in Fig.~\ref{fig:df_dz_scatter}.

The enhanced decay of axion DM induced by soliton mergers may have phenomenological consequences that either offer new windows to axion indirect detection, or place stronger constraints on the axion parameter space. Soliton decay injects energy into the intergalactic medium, which, for the reference parameters shown in Fig.~\ref{fig:energy}, is significantly larger than the energy injection due to core collapse supernovae, and extends to much higher redshifts. We explore phenomenology of this energy injection in detail in a companion paper~\cite{Escudero:2023vgv}.

\section*{Acknowledgements}

D.J.E.M. is supported by an Ernest Rutherford Fellowship from the Science and Technologies Facilities Council (No. ST/T004037/1). D.J.E.M. thanks Tilman Hartwig and Mattis Magg for useful correspondence on supernova energy injection.
D.B. is supported by a ``Ayuda Beatriz Galindo Senior" from the Spanish ``Ministerio de Universidades," grant No. BG20/00228.  M.F. and C.K.P. are funded by the Science and Technology Facilities Council (UK).
The research leading to these results has received funding from the Spanish Ministry of Science and Innovation (No. PID2020-115845GB-I00/AEI/10.13039/501100011033).
IFAE is partially funded by the Centres de Recerca de Catalunya (CERCA) program of the Generalitat de Catalunya. D.B. acknowledges the support from the Departament de Recerca i Universitats de la Generalitat de Catalunya al Grup de Recerca i Universitats from Generalitat de Catalunya to the Grup de Recerca 00649 (Codi: 2021 SGR 00649).

Computing resources used in this work were made available by a generous grant from the Ahmanson Foundation.


\appendix

\section{Details of the Calculation}\label{appendix:calc_details}

Our calculation of the soliton merger rate at very high redshift and at very low halo mass requires the computation of many nested integrals covering a wide range of scales. We describe briefly here how this calculation is done in a numerically efficient and physically accurate manner.

We used \textsc{class}~\cite{2011arXiv1104.2932L} to compute the matter power spectrum, $P(k)$, in a standard CDM cosmology. We then applied the analytic approximation for the effect of the Jeans scale cut-off given by Ref.~\cite{Hu:2000ke}. We chose this methodology, rather than direct computation of the Jeans scale using \textsc{axionCAMB}~\cite{Hlozek:2014lca} because: (a) we use \textsc{class} in Ref.~\cite{Escudero:2023vgv} to compute the reionization effect of soliton decays, and (b) the analytic approximation is simpler to implement at larger particle mass $m_a$ and wave number $k$ of interest at present.~\footnote{See Ref.~\cite{Passaglia:2022bcr} for discussion of the accuracy of these various methods for computing the Jeans scale cutoff.}  

We pre-compute $\sigma(M)$ at $z=0$ at fixed particle masses, $m_a$ and fixed cosmological parameters, $\Omega_m = 0.3153$, $\Omega_b=0.04930$, $\Omega_\Lambda  = 0.6847$, $\sigma_8 =0.8111$, $n_s=0.9649$, $h=0.6736$~\cite{Planck:2018jri}. We first compute $\sigma(R)$ using three different window functions, $\widetilde{W}(k|R)$ discussed in the text: (a) real space top-hat, (b) sharp-$k$ space, (c) smooth-$k$ space window function~\cite{Leo:2018odn,Bohr:2021bdm}. These are given by:
\begin{align}
    \widetilde{W}^{\text{top-hat}}(k|R)&= \frac{3\left[\sin(kR)-kR\cos(kR)\right]}{(kR)^3}, \\
    \widetilde{W}^{\text{sharp-k}}(k|R)&= \theta\left(1-kR\right), \\
    \widetilde{W}^{\text{smooth-k}}(k|R)&= \frac{1}{1+\left(kR\right)^{\beta}}.
    \label{eqn:smooth_k_window} 
\end{align}
Here $\theta$ is the Heaviside step function. The variance is defined by:
\begin{equation}
    \sigma^2(R) =  \int_0^{\infty}\frac{k^2}{2\pi^2}P(k)\widetilde{W}(k|R)^2 \mathrm{d}k\, .
\end{equation}

In the calculation of the soliton merger rate using the EPS formalism in Sec.~\ref{sec:eps_merger}, we adopt the smooth-$k$ space window function, Eq.~\eqref{eqn:smooth_k_window} as a reference, with mass assignment Eq.~\eqref{eqn:mass_assign}. This is because this mass variance reproduces the HMF cutoff seen in simulations very well directly from $\sigma(M)$, without the need for additional fitting functions. This $\sigma(M)$ can then be used as direct input in the merger rate calculation.

To find the mass variance, we need to assign mass to a scale $(R)$. This is trivial for the real-space top hat, but requires calibration for both sharp-$k$ and smooth-$k$ space window functions. We use the following mass assignments:
\begin{align}
M^{\text{top-hat}}(R)&= \frac{4}{3}\pi R^3 \rho_m, \\
M^{\text{sharp-k}}(R)&= \frac{4}{3}\pi (a_W R)^3 \rho_m, \\
M^{\text{smooth-k}}(R)&= \frac{4}{3}\pi (c_W R)^3 \rho_m.
\label{eqn:mass_assign}
\end{align}
For the sharp-$k$ filter we take $a_W=2.5$ as in~\cite{Benson:2012su}. For the smooth-$k$ filter, we take $c_W=2.15940$ and $\beta=9.10049$ which were found by fitting large $N$-body simulations~\cite{Benson2023}.

A converged computation of $\sigma(M)$ at the required low values of $M$ requires computing $P(k)$ to large $k$. In \textsc{class} we set $k_{\rm max} = 2.5\times 10^{4}\text{ Mpc}^{-1}$, and extrapolate $P(k)$ at $k > k_{\rm max}$ using a power law. The large value of $k_{\rm max}$ required might suggest that an analytic approximation to $P(k)$ would be useful. However, the best such approximation, Ref.~\cite{Eisenstein:1997ik} is not accurate on $k$ larger than the baryon Jeans scale, and gives $P(k)$ which is too large compared to the one computed accurately by \textsc{class} direct solution of the Boltzmann equation.

For cosmological parameters adopted in this paper, we found the following fitting functions for $\sigma(M)$ for both CDM and axion DM models:
\begin{eqnarray}
\sigma^{\rm CDM}(M)&=&a \left(\frac{M}{10^{3}M_{\odot}}\right)^{-b} \nonumber \\
&&\left[1-c \ln \left(1+d \sqrt{\frac{M}{10^{3}M_{\odot}}}\right)\right],
\label{eq:sigma_fit_CDM}\\
\sigma^{\rm axion}(M)&=&\left[1+\left(\frac{M}{e M_{\rm half}}\right)^f\right]^{b/f}\sigma^{\rm CDM}(M),
\label{eq:sigma_fit_axion}
\end{eqnarray}
where $M_{\rm half}$ is the half-mode mass
\begin{eqnarray}
M_{\rm half} &=& \frac{4}{3}\pi \left(\frac{\pi}{k_{\rm half}}\right)^3 \rho_m,
\label{eq:M_half_mode} \\
k_{\rm half} &=&  4.986\times 10^4 \left(\frac{m_a}{10^{-13}{\rm eV}}\right)^{4/9}{\rm Mpc}^{-1},
\label{eq:k_half_mode}
\end{eqnarray}
and the best-fit parameters take
\begin{eqnarray}
a&=&11.2934,\\
b&=&0.0231895,\\
c&=&0.0800510,\\
d&=&0.125902,\\
e&=&0.329159,\\
f&=&-2.41133.
\end{eqnarray}
Note that the fitting functions above are only fitted to halo masses smaller than $10^{12} M_{\odot}$ and axion masses larger than $10^{-17} {\rm eV}$. For those cases, they provide an accuracy better than $1\%$. To compute the halo mass function and halo formation rate, one also need to compute $\alpha \equiv \mathrm{d}\ln \sigma/\mathrm{d}\ln M$.~\footnote{It should not be confused with the slope parameter in the core-halo mass relation Eq.~\eqref{eqn:core-halo}.} This can easily be done using the fitting functions Eqs.~\eqref{eq:sigma_fit_CDM} and~\eqref{eq:sigma_fit_axion}. But we found that directly computing $\alpha$ for axion DM models using Eq.~\eqref{eq:sigma_fit_axion} leads to large errors for $M\ll M_{\rm half}$. So we provide an additional fitting function for $\alpha^{\rm axion}$ below.
\begin{equation}
\alpha^{\rm axion}(M)=\left[1+\left(\frac{M}{e M_{\rm half}}\right)^f\right]^{g/f}\alpha^{\rm CDM}(M)\,.
\label{eq:alpha_fit_axion}
\end{equation}
Here $\alpha^{\rm CDM}(M)\equiv\mathrm{d}\ln \sigma^{\rm CDM}/\mathrm{d}\ln M$ is computed using Eq.~\eqref{eq:sigma_fit_CDM} and $g=2.86571$.

With the above fitting functions, the halo mass function can be computed as
\begin{equation}
\frac{\mathrm{d}n_{\rm h}}{\mathrm{d} M} = 2\frac{\rho_m}{M} \frac{\mathrm{d} f^{\rm ST}(S(M),z)}{\mathrm{d} S}\frac{\sigma(M)^2}{M}|\alpha(M)|\,,
\label{eq:HMF}
\end{equation}
where $\mathrm{d} f^{\rm ST}(S,z)/\mathrm{d} S$ is given by Eq.~\eqref{eq:f_ST}.

\section{Generating Merger Trees}\label{appendix:merger_tree}

\begin{figure}[t]
\begin{center}
\includegraphics[width=\columnwidth]{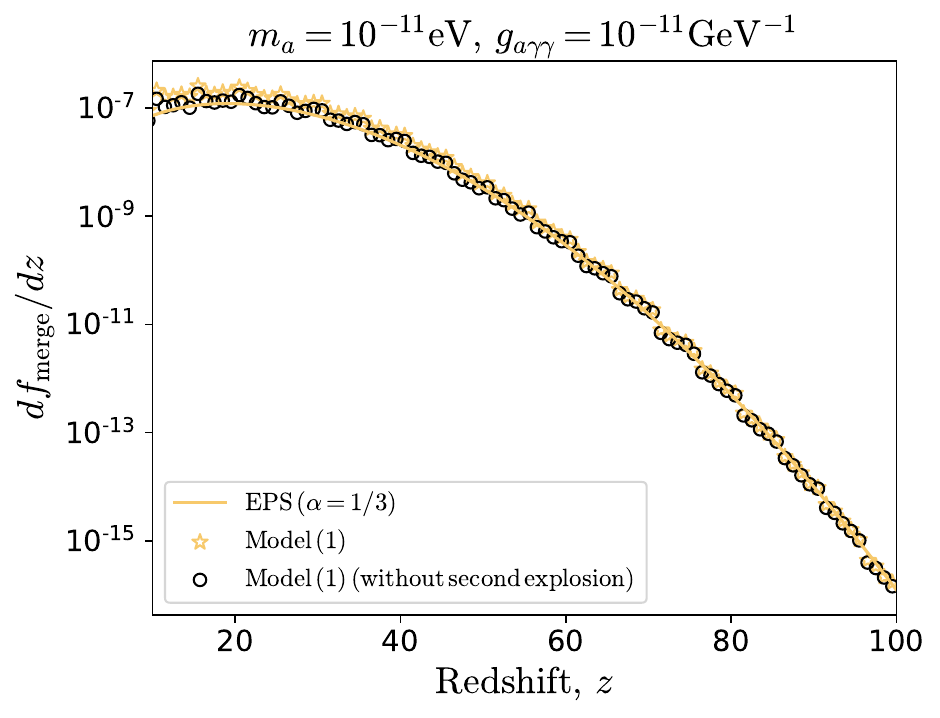}
\caption{DM fractional decay rate from merger trees assuming core growth model from Ref.~\cite{Du:2016aik}. If a second explosion is allowed, the result is slightly higher at lower redshifts.
}
\label{fig:df_dz_du_comapre}
\end{center}
\end{figure}

The merger trees are generated using the semi-analytic code \textsc{Galacticus}~\cite{Benson_2012} which employs the EPS Monte Carlo algorithm proposed in Ref.~\cite{Parkinson:2007yh} (see also Ref.~\cite{Cole:2000ex}). To compute the formation rate of axion stars with critical masses at different redshifts, we start building the merger from a final redshift of $z_f$. As we go backward in time along the merger trees, halos fragment into smaller and smaller progenitor halos. After significant fragmentation has occurred the number of halos with masses in the range we are interested in becomes too small, so we only use the data in the redshift range $[z_f, z_f+5]$. At this point a new set of trees is built starting from from $z_f+5$. For each case, we generate $400,000$ trees with root halo masses sampled from the halo mass function in the mass range $[10^{-7},10^3]\,{M}_{\odot}$. The mass resolution of the tree is set to $10^{-7}\,{M}_{\odot}$ to make sure that for the cases we shown in Sec.~\ref{sec:merger_trees}, the critical halo mass is always resolved. For $m_a=10^{-11}{\rm eV}$, $g_{a\gamma\gamma}=10^{-12} {\rm GeV}^{-1}$, and the core-halo mass relation from Ref.~\cite{Schive:2014dra}, the critical halo mass $M_{\rm h,crit}$ at $z=10$ ($z=100$) is $11.4 M_{\odot}$ ($0.4 M_{\odot}$). If we sample the root halo masses from the halo mass function, we will have too few halos close to $M_{\rm h,crit}$ making the calculation of formation rate inaccurate. So for this case, we sample the root halo masses from a loguniform distribution in the mass range $[10^{-2},10^{3}]\,{M}_{\odot}$ instead. To get the correct formation rate, each tree is assigned a weight
\begin{equation}
w_i=\int_{M_{i-1/2}}^{M_{i+{1/2}}} \frac{\mathrm{d}n}{\mathrm{d}M} \mathrm{d}M,
\label{eq:tree_weight}
\end{equation}
where $\{M_i\}$ is in ascending order and $M_{i-1/2}=\sqrt{M_i M_{i-1}}$.

Having the merger trees, we can assign each halo in the tree a core mass $M_\mathrm{c}$ (the axion star mass $M_S\approx4 M_\mathrm{c}$) using either a chosen core-halo mass relation or the model proposed in Ref.~\cite{Du:2016aik} (see Sec.~\ref{sec:merger_trees}). For the latter case, we include some additional physics: (1) when the axion star in the center of a halo is above the critical star mass $M_{\rm S,crit}$, we set $M_\mathrm{c}=0$; (2) if a halo, with mass above $M_{\rm h,crit}$, and whose central axion star has completely decayed, has another major merger with a halo below $M_{\rm h,crit}$, the axion star from the second progenitor will remain in the new halo, i.e. it is not disrupted (see Fig.~\ref{fig:merger_tree}). The major merger mass ratio is defined as $1:13$.~\footnote{The result is only slightly affected by the specific choice of this ratio.} With the second assumption, a halo can have a second explosion of the central axion star, which leads to slightly higher dark matter decay rate at lower redshifts, see Fig.~\ref{fig:df_dz_du_comapre}.

\section{Timescale of merging process}\label{appendix:time_scale}

In Sec.~\ref{sec:merger_rate}, when we compute the decay rate of axion stars due to major mergers, we have assumed that the axion stars in the center of halos merge at the same time when two halos merge. In reality, when two halos encounter each other, the smaller one will become the subhalo of the larger one (the host). The orbit of the subhalo decays due to dynamical friction~\cite{Chandrasekhar1943}. As its orbit decays with time, the subhalo gradually sinks into the host's center. The central axion star of the subhalo finally merges with the one residing in the host's center if it is not disrupted by tidal forces. The dynamical friction is stronger for more massive subhalos, thus the timescale of the merging process, $\tau_{\rm merging}$, is shorter for larger mass ratios $\mu_{\rm h}=M_{\rm sub}/M_{\rm host}$.

To obtain the timescale $\tau_{\rm merging}$, we simulate the merger of two halos with different mass ratios using the $N$-body code Gadget-4~\cite{Springel:2020plp}. In this work, we are only concerned with major mergers, i.e. $\mu_{\rm h} \geq (3/7)^{1/\alpha}$, which lead to the growth of axion stars (for more details, see the discussions in Sec. \ref{sec:DM_decay_rate}).

As an example, we consider a host mass of $10^{-3}M_{\odot}$ which corresponds to a halo containing an axion star close to the critical mass at $z=20$ for $m_a=10^{-11}{\rm eV}$, $g_{a\gamma\gamma}=10^{-11}{\rm GeV}^{-1}$, and $\alpha=1/3$ (assuming a core-halo mass relation found in Ref.~\cite{Schive:2014hza}). The subhalo first enters the host's virial radius, $R_{\rm vir, host}$, at redshift $z_{\rm infall}=20$, at which redshift both halos are assumed to have Navarro–Frenk–White (NFW) profiles~\cite{Navarro:1995iw}. The concentration parameters of the NFW profiles are computed using the model proposed in Ref.~\cite{Diemer:2018vmz} \footnote{The model in Ref.~\cite{Diemer:2018vmz} is shown to work reasonably well for the halo mass we consider here. At $z=0$ the concentration predicted by this model for a halo of mass $10^{-3} M_{\odot}$ is only slightly higher than that found in numerical simulations~\cite{Wang:2019ftp}.}. We start the simulation at $z_{\rm infall}$ and consider two types of initial orbits: (1) the subhalo is initially on a circular orbit with a zero radial velocity and a tangential velocity computed as $V_t=\sqrt{G M_{\rm host}/R_{\rm vir, host}}(1+\mu_{\rm h})$; (2) the subhalo has typical radial and tangential velocities found in cosmological simulations~\cite{Jiang:2014zfa}.

\begin{figure*}[t]
\begin{center}
\includegraphics[width=\columnwidth]{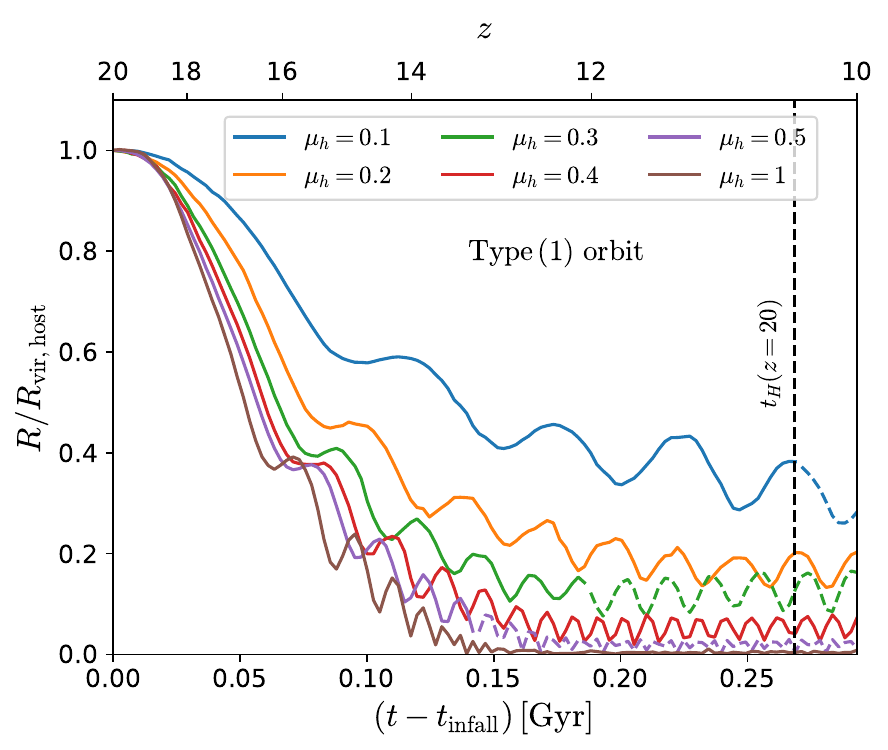}
\includegraphics[width=\columnwidth]{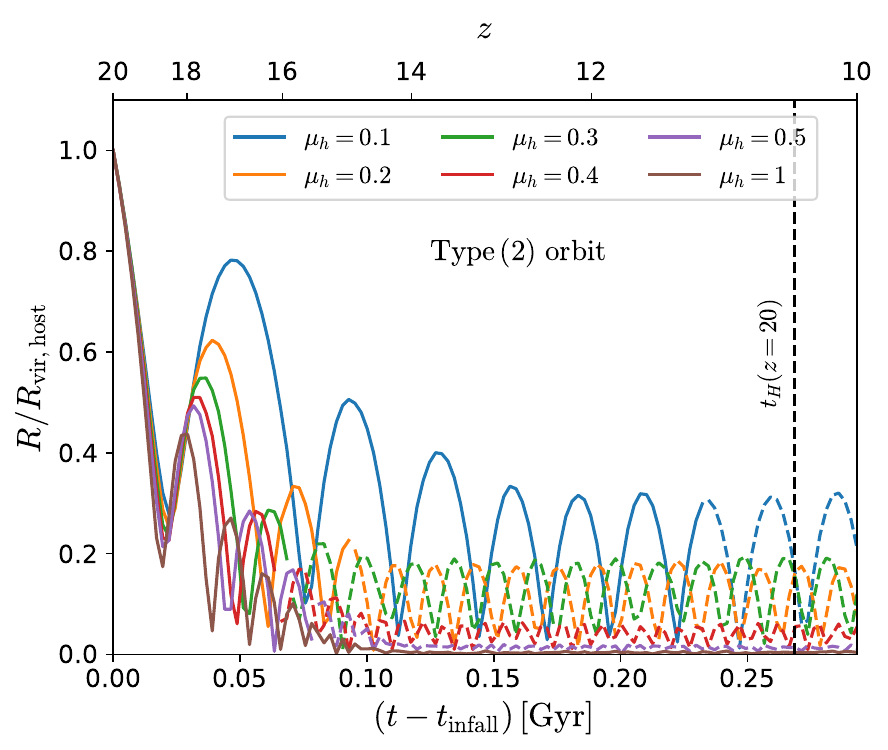}
\caption{Orbital decays caused by dynamical friction for halos mergers with different mass ratios. The host has a mass of $10^{-3} M_{\odot}$. The subhalo falls into the host's virial radius, $R_{\rm vir,host}$, at $z_{\rm infall}=20$. \emph{Left:} type (1) orbit, the subhalo is initially on a circular orbit. \emph{Right:} type (2) orbit, the subhalo is initially on an elliptical orbit with the radial and tangential velocity taken for typical infalling subhalos~\cite{Jiang:2014zfa}. When the subhalo contains less than $20$ bound particles, its position tracked by the most-bound particle is show as dashed curve.
}
\label{fig:orbit_decay}
\end{center}
\end{figure*}

Figure~\ref{fig:orbit_decay} shows the distance between the subhalo and the host as a function of time since infall obtained from our simulations. The results for type (1) and type (2) initial orbits are shown in the left and right panels, respectively. Compared to type (1) orbits, subhalos with type (2) orbits can reach further inside the host where the host density is higher, thus are subjected to stronger dynamical friction and have smaller $\tau_{\rm merging}$. For $\mu_{\rm h}=0.1$, the orbit of the subhalo decays rapidly at early times, but then the subhalo remains on a nearly stable orbit when the dynamical friction becomes negligible. Instead of sinking into the host's center, the subhalo gradually loses its mass and may finally be destroyed by the tidal forces from the host. When the number of particles gravitationally bound to the subhalo drops below $20$, we track the position of the subhalo by the most-bound particle, the particle that has the most negative potential energy (identified in the last snapshot at which the subhalo contained more than $20$ bound particles). The orbits of the most-bound particle are shown as dashed curves. For $\mu_{\rm h}\gtrsim 0.3$, the subhalo is able to reach close enough to the host's center so that the axion star in the subhalo can merge with that in the host. 

The $N$-body simulations we perform cannot describe the axion stars in the halos and their mergers. Thus we only obtain a rough estimate of $\tau_{\rm merging}$ from Fig.~\ref{fig:orbit_decay}. While type (2) orbits are better representations of realistic subhalo orbits, we take $\tau_{\rm merging}$ from the type (1) simulations as a conservative estimate. For $\mu_{\rm h}\gtrsim 0.3$, we find $\tau_{\rm merging}\sim 0.16{\rm Gyr}$. After the merger of two axion stars, if the new axion star has a mass above the critical value given by Eq.~\eqref{eqn:decay-mass}, the axion particles in the axion star will decay to photons exponentially within a timescale negligible compared to $\tau_{\rm merging}$. So we expect the axion decays are delayed by $\tau_{\rm merging}$ compared to the time that we calculate in Sec.~\ref{sec:merger_rate}. Note that $\tau_{\rm merging}$ is smaller than the Hubble time at $z_{\rm infall}$ (vertical dashed curve in Fig.~\ref{fig:orbit_decay}), so the effect of delay is expected to be minor. Simulations that are capable of correctly including the quantum pressure of the scalar field are needed to study the merging process more accurately. We leave such a detailed and statistical analysis to further work (although see Ref.~\cite{Du:2018qor}).  

\begin{figure}[t]
\begin{center}
\includegraphics[width=\columnwidth]{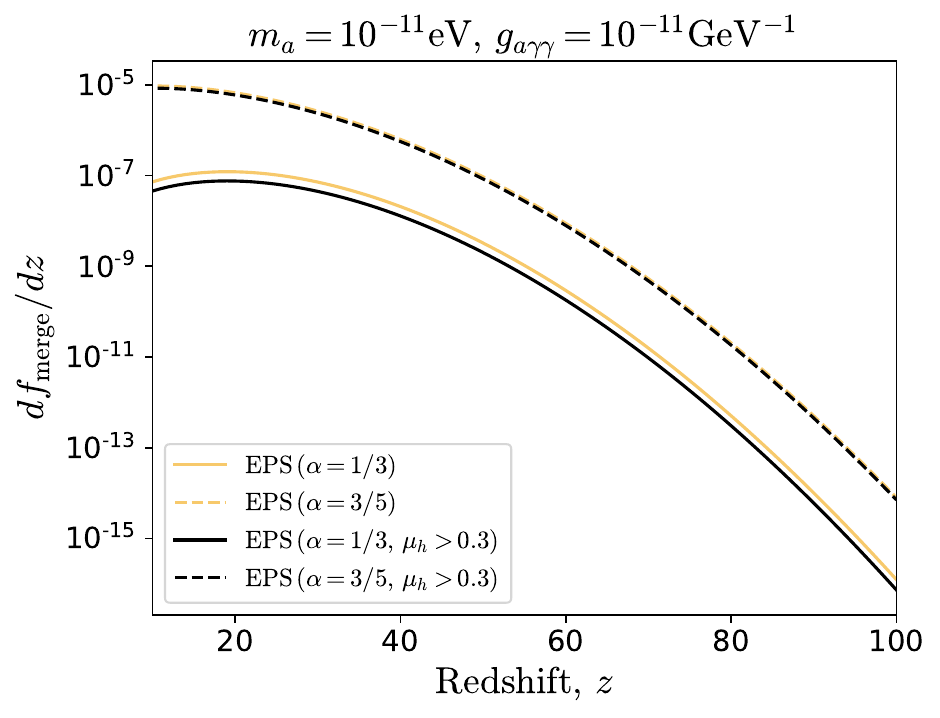}
\caption{DM fractional decay rate due to soliton mergers with $\mu_{\rm h}>0.3$ (black curves) comparing with our fiducial EPS models (colored curves).}
\label{fig:df_dz_mu_cut}
\end{center}
\end{figure}

As shown above, not all major mergers as we defined in Sec.~\ref{sec:merger_rate}, i.e., $\mu_{\rm h} \geq (3/7)^{1/\alpha}$, result in the merger of the central axion stars in halos. To check how this may affect the axion decay rate, we impose an additional condition, $\mu_{\rm h}\geq 0.3$, when computing the integral Eq.~\eqref{eq:formation_rate_major}. Figure~\ref{fig:df_dz_mu_cut} shows the DM fractional decay rate due to mergers if we impose this additional condition comparing with that from our fiducial EPS models. For $\alpha=1/3$ ($\alpha=3/5$), the DM fractional decay rate is reduced by $40\%$ ($10\%$), a relatively small effect on the scales shown that does not affect our main conclusions.

\newpage

\bibliography{biblio}

\end{document}